%% file: main.tex
\documentclass[sigconf, authorversion]{acmart}

\usepackage{epsfig,endnotes}
\pdfoutput=1
\usepackage{enumitem}
\usepackage{graphicx}
\usepackage{enumitem}
\usepackage{epstopdf}
\usepackage{amssymb}
\usepackage{amsmath,amsthm}
\usepackage{comment}
\usepackage{graphicx} 
\usepackage{longtable}
\usepackage{multirow}
\usepackage{booktabs}
\usepackage{tabularx}
\usepackage{algorithm2e}
\usepackage{makecell}
\usepackage{blindtext}
\usepackage{url}
\usepackage{listings}
\usepackage{subfig}
\usepackage{color}
\usepackage{fancyhdr}
\usepackage{textcomp}
\usepackage{gensymb}
 \usepackage{enumitem}
\usepackage{makecell}
\usepackage{array}
\usepackage{colortbl}
\newcolumntype{P}[1]{>{\centering\arraybackslash}p{#1}}
\usepackage{caption}
\usepackage[flushleft]{threeparttable}
\definecolor{Gray}{gray}{0.9}
\usepackage{appendix}
\usepackage{stfloats}
\usepackage{float}
\usepackage[utf8]{inputenc}

\usepackage{todonotes}

\usepackage{wrapfig}
\usepackage{mathtools}
\newcommand{\projectname}{{\textsc{\small{Kratos}}}\xspace}

\usepackage{tikz}

\usepackage{amsmath}
\iftrue
\newcommand{\berkay}[1]{{\color{red}{\bf BC:} #1}}
\else
\newcommand{\berkay}[1]{}
\fi

\usepackage{caption}
\usepackage[flushleft]{threeparttable}
\definecolor{Gray}{gray}{0.9}
\usepackage{appendix}
\usepackage{stfloats}
\usepackage{float}
\usepackage{color}
\definecolor{light-gray}{gray}{0.90}
 \makeatletter
\newcommand{\algorithmfootnote}[2][\footnotesize]{%
 \let\old@algocf@finish\@algocf@finish
 \def\@algocf@finish{\old@algocf@finish
    \leavevmode\rlap{\begin{minipage}{\linewidth}
    #1#2
    \end{minipage}}%
  }%
}
\makeatother
\usepackage{geometry}
\usepackage[utf8]{inputenc}
\graphicspath{ {WiSec 2020/images/} }
\usepackage{wrapfig}
\usepackage{mathtools}

\usepackage{tikz}

\usepackage{amsmath}

\makeatletter 
\newcommand\semiHuge{\@setfontsize\semiHuge{22.72}{27.38}}
\makeatother 

\usepackage{listings}
\usepackage{color}
\DeclareCaptionFont{black}{\color{black}}
\usepackage{tikz}
\usetikzlibrary{shapes.geometric,calc}

\newcommand\score[2]{

\pgfmathsetmacro\pgfxa{#1+1}

\tikzstyle{scorestars}=[star, star points=5, star point ratio=2.25, draw, inner sep=0.15em,anchor=outer point 3]

\begin{tikzpicture}[baseline]
  \foreach \i in {1,...,#2} {
    \pgfmathparse{(\i<=#1?"black!70":"lightgray")}
    \edef\starcolor{\pgfmathresult}
    \draw (\i*1em,0) node[name=star\i,scorestars,fill=\starcolor]  {};
   }
  \pgfmathparse{(#1>int(#1)?int(#1+1):0}
  \let\partstar=\pgfmathresult
  \ifnum\partstar>0
    \pgfmathsetmacro\starpart{#1-(int(#1))}
    \path [clip] ($(star\partstar.outer point 3)!(star\partstar.outer  point  2)!(star\partstar.outer point 4)$) rectangle 
    ($(star\partstar.outer point 2 |- star\partstar.outer point   1)!\starpart!(star\partstar.outer point 1 -| star\partstar.outer point 5)$);
    \fill (\partstar*1em,0) node[scorestars,fill=black!70]  {};
  \fi,

\end{tikzpicture}
}
\newcommand{\bcircle}{
\begin{tikzpicture}
\filldraw[fill=black,draw=black] circle (2pt);
\end{tikzpicture}
}

\newcommand{\wcircle}{
\begin{tikzpicture}
\filldraw[fill=white,draw=black] circle (2pt);
\end{tikzpicture}
}

\copyrightyear{2020}
\acmYear{2020}
\setcopyright{acmcopyright}\acmConference[WiSec '20]{13th ACM Conference on Security and Privacy in Wireless and Mobile Networks}{July 8--10, 2020}{Linz (Virtual Event), Austria}
\acmConference[WiSec '20]{13th ACM Conference on Security and Privacy in Wireless and Mobile Networks}{July 8--10, 2020}{Linz (Virtual Event), Austria}
\acmBooktitle{13th ACM Conference on Security and Privacy in Wireless and Mobile Networks (WiSec '20), July 8--10, 2020, Linz (Virtual Event), Austria}
\acmPrice{15.00}
\acmDOI{10.1145/3395351.3399358}
\acmISBN{978-1-4503-8006-5/20/07}

\begin{document}
\title{\textsc{Kratos}: Multi-User Multi-Device-Aware Access Control System for the Smart Home} 
\author{Amit Kumar Sikder$^1$, Leonardo Babun$^1$, Z. Berkay Celik$^2$, Abbas Acar$^1$, Hidayet Aksu$^1$, Patrick McDaniel$^3$, Engin Kirda$^4$, A. Selcuk Uluagac$^1$}

\affiliation{%
 \institution{$^1$Florida International University -  \{asikd003, lbabu002, aacar001, haksu, suluagac\}@fiu.edu}
}

\affiliation{%
 \institution{$^2$Purdue University -  zcelik@purdue.edu}
}

\affiliation{%
 \institution{$^3$Penn State University - mcdaniel@cse.psu.edu}
}

\affiliation{%
 \institution{$^4$Northeastern University - ek@ccs.neu.edu}
}
\renewcommand{\shortauthors}{Sikder et al.}

\input{abstract.tex}



\maketitle

\input{introduction.tex}
\input{background.tex}
\input{problem.tex}

\input{technical.tex}

\input{implementation.tex}

\input{evaluation.tex}

\input{benefits.tex}

\input{relatedwork.tex}
\input{conclusion.tex}
\vspace{-0.1cm}
\section{Acknowledgment}\label{sec:acknowledgment}
This work is partially supported by the US National Science Foundation (Awards: NSF-CAREER-CNS-1453647, NSF-1663051, NSF-1705135), US Office of Naval Research grant Cyberphysical Systems, and Cyber Florida’s Capacity Building Program. The views expressed are those of the authors only, not of the funding agencies.

{\footnotesize\bibliographystyle{ACM-Reference-Format}
\bibliography{Bibtex.bib}}

\input{appendix1}

\end{document}

%% file: abstract.tex
\begin{abstract}
In a smart home system, multiple users have access to multiple devices, typically through a dedicated app installed on a mobile device. Traditional access control mechanisms consider one unique trusted user that controls the access to the devices. However, multi-user multi-device smart home settings pose fundamentally different challenges to traditional single-user systems. For instance, in a multi-user environment, users have conflicting, complex, and dynamically changing demands on multiple devices, which cannot be handled by traditional access control techniques. To address these challenges, in this paper, we introduce \projectname, a novel multi-user and multi-device-aware access control mechanism that allows smart home users to flexibly specify their access control demands. \projectname has three main components: user interaction module, backend server, and policy manager. Users can specify their desired access control settings using the interaction module which are translated into access control policies in the backend server. The policy manager analyzes these policies and initiates negotiation between users to resolve conflicting demands and generates final policies. We implemented \projectname and evaluated its performance on real smart home deployments featuring multi-user scenarios with a rich set of configurations (309 different policies including 213 demand conflicts and 24 restriction policies). These configurations included five different threats associated with access control mechanisms. Our extensive evaluations show that \projectname is very effective in resolving conflicting access control demands with minimal overhead, and robust against different attacks.
\end{abstract}

%% file: introduction.tex
\vspace{-0.1in}
\section{Introduction}\label{sec:intro}
Cyberspace is expanding fast with the introduction of new smart home technologies dedicated to make our homes automated and smarter~\cite{forbes2, sikder2018iot}. This trend will only continue, and billions of smart devices will dominate our everyday lives by the end of this decade~\cite{business-insider, sikder2019aegis}. The smart home systems (SHSs) allow multiple devices to be connected to automate daily activities and increase the overall efficiency of the home. Devices as simple as a light bulb to ones as complicated as an entire AC system can be connected and exposed to multiple users. The users then interact with the devices through different smart home applications installed through a mobile host app provided by the smart home vendors. Traditional access control mechanisms proposed for personal devices such as computers and smartphones primarily target single-user scenarios. However, in a SHS, multiple users access the same smart device, typically via a common controller app (e.g., SmartThings App), which can cause conflicting device settings. For instance, a homeowner may want to lock the smart door lock at midnight while a temporary guest may want to access the lock after midnight. Also, current smart home platforms do not allow the conflicting demands of the users to be expressed explicitly. Finally, the current access control mechanism in smart home platforms offer coarse-grained solutions that might cause safety and security issues~\cite{Berkay2018Soteria, jia2017contexiot, babun2018iotdots}. For instance, smart home platforms often give automatic full access to every user added to the SHS~\cite{access}. With full access, a new user can easily add new unauthorized users, remove existing users, or reconfigure the connected devices~\cite{203866, guan2017smart}. This benign, yet undesired action from the new user can lead to several safety issues~~\cite{sikder2018survey, spmagazine, newaz2019healthguard}. In these real-life scenarios, current smart home platforms cannot fulfill such complex, asymmetric, and conflicting demands of the users as they can only handle primitive and broad controls with static configurations.

In this paper, we introduce \projectname, a multi-user multi-device-aware access control system designed for the SHSs. \projectname introduces a formal policy language that allows users to define different policies for smart home devices, specifying their needs. It also implements a policy negotiation algorithm that automatically solves and optimizes the conflicting policy requests from multiple users by leveraging user roles and priorities. Lastly, \projectname governs different policies for different users, reviewing the results of the policy negotiation and enforcing the negotiation results over the smart home devices and apps. We implemented \projectname in a real a multi-user multi-device SHS that include 17 different sensors and actuators. We further evaluated \projectname performance on 219 different policies including 146 demand conflicts and 33 restriction policies collected from real-life smart home users. We also assessed the performance of \projectname against five different threat models. Our extensive evaluation shows that, \projectname can resolve demand conflicts and detect different threats with 100\% success rate in a multi-user multi-device SHS with minimal overhead.

\vspace{3pt}\noindent \textbf{Contributions:} The main contributions of this work are as follows:

\begin{itemize}[wide=0pt]

\item We introduced \projectname, a multi-user multi-device-aware access control system for SHS. \projectname provides a flexible policy-based user controls to define user roles and understand users' demands on smart home, a formal policy language to express users' desires, and a policy negotiation mechanism to automatically resolve and optimize conflicting demands and restrictions in a multi-user SHS.

\item We implemented \projectname on a real SHS using 17 different smart home devices and sensors. Further, we evaluated its performance with 309 different policies provided by real users. Our evaluation results show that \projectname effectively resolves conflicting demands with minimal overhead.

\item We tested \projectname against five different threats arising from inadequate access control system. Our evaluation shows that \projectname can detect different threats with 100\% success rate.

\end{itemize}

\vspace{1pt}\noindent \textbf{Organization:} In section~\ref{background}, we present the needs of access control in SHS. In Section~\ref{problem}, we articulate the problem space and explain the threat model. We detail the design of \projectname in Section~\ref{sec:technical}. Section~\ref{sec:implement} articulates the implementation of \projectname in a real-life setting. In Section~\ref{sec:evaluation}, we evaluate the performance of \projectname. Finally, Section~\ref{sec:related} discusses the related work and Section~\ref{sec:conclusion} concludes the paper. 

%% file: background.tex
\section{Motivation and Definitions}\label{background}

\subsection{Access Control Needs in a Smart Home}
Access control in multi-user SHSs pose unique challenges in terms of device sharing and conflict resolution. People sharing smart devices in the same environment may have different needs and usage patterns which can lead to conflict scenarios~\cite{205156}. However, existing smart home platforms mostly offer binary control mode where a user gets all the control or no control at all. For instance, \textit{Samsung SmartThings} provides full access to all the connected devices to a user. Unfortunately, this all-inclusive access permits any authorized users to control the smart devices which can lead to conflicting demands, privacy violations, and undesired app installations~\cite{geeng2019s}. To protect smart devices from unauthorized app installation and device settings, some smart home platforms (e.g., \textit{Apple HomeKit}) offers two device access options: remote access, and editing. In remote access, a new user gets access only privilege to the connected devices. In the editing option, a new user obtains permission for adding or removing any app, device, or user in the system. Additionally, some smart home devices (e.g., August smart lock, Kwikset Kevo Smart Lock, etc.) offers temporary user access or guest access to limit undesired access after an expiry time~\cite{lock, remote}. These solutions, however, are vendor- and device-specific, thus, are not ready and applicable in a multi-device multi-user smart home system. In summary, \textit{existing access control mechanism in smart home technologies fail to deliver the diverse and complex user demands in a multi-device multi-user setting.}

As conflicting scenarios in a multi-user SHS depend on users' relationships, social norms, and personal preferences, it is important to understand these dynamics before designing a fine-grained access control system. For instance, parents may want to restrict smart TV access for the kids, roommates may want privacy for bedroom locks in a sharing apartment, or owner may want to give temporary access to Airbnb guests. Hence, a fine-grained access control system should address diverse needs of the users in a multi-user multi-device smart home ecosystem. Several prior works have focused on understanding the user preferences and needs by conducting user studies among smart home users~\cite{217501, zeng2019understanding, zeng_soups}. These prior works have established the needs and design requirements of access control systems in SHSs. He et al. conducted a user study among 425 smart home users to understand how relationships among the users affect the access control needs in a SHS~\cite{217501}. In a recent work, Zeng et al. developed an early prototype of access control system and conduct a usability study among 8 households to identify the access control needs in a real-life SHS~\cite{zeng2019understanding}. These user studies reported the following preferences and needs among smart home users in terms of access control.

\begin{itemize}[wide=0pt]
    \item Majority of the users expressed the need of a fine-grained access control system in a SHSs. 
    \item Users suggested role-based access control (RBAC) in a home environment for limiting access to the devices and applications.
    \item In a shared environment, users agreed for per-device roles for private rooms.
    \item For temporary users in a shared environment, users suggested location-based and time-based access control.
    \item For conflicting scenarios, users expressed the need of automation rules to resolve the conflicting demands and configure the devices at an optimum operational value.
\end{itemize}
In our work, we consider the users' needs and suggestions reported in prior works to design a fine-grained access control system for multi-user multi-device SHSs.

\subsection{Terminology}\label{defn}
We define several important terms that we use in this work.

\vspace{2pt}\noindent\textbf{Policy.} We consider \textit{Policy} as the group of requests made by the users to control device usage in a multi-user smart environment. Based on the nature of request, there are three types of policies.  
\begin{enumerate}[wide=0pt]
    \item\textit{Demand Policy.} We consider \textit{Demand Policy} as the group of requests made by a user that define the control rules for a specific device or group of devices in the smart home system. Demand policies can be general (i.e., created by the admin and applied to all the users in the system) or specific to a certain user. If a demand policy is general to all users, we define that as \textit{General Policy}.
    \item\textit{Restriction Policy.} We consider \textit{Restriction Policy} as the set of rules that govern the accessibility and level of control of a user or group of users to a certain device or group or devices in the smart home system. Restriction policies regulate (1) what devices the user has access to, (2) the time frame in which the user is authorized to use/control the device, and (3) the control setting limits.
    \item\textit{Location-based Policy} We consider \textit{Location-based Policy} as the set of automation rules enforced by the user that are only applicable if the user is connected in the system network. Location-specific policies regulate (1) what devices the user has remote access to and (2) the control setting limit if a specific user is not present in the smart home network. 
\end{enumerate}

\noindent\textbf{Priority.} We call \textit{Priority} as the importance level of a user that may be used to create preferences for users of higher priority over users with lower priority during new user addition, restriction, and demand negotiation processes. In Section~\ref{sec:technical}, we detail the different priority levels considered in this work. 

\noindent\textbf{Conflict.} For the purpose of this work, \textit{Conflict} is defined as the dispute process that is generated from two or more demand policies that interfere or contradict based on the specific requests of the policies. Based on the nature of the demand and restriction policies, three types of conflict can occur.
\begin{enumerate}[wide=0pt]
    \item \textit{Hard conflict.} A hard conflict occurs when demand policies of a specific device enforced by two different users do not have any overlapping device condition. 
    \item \textit{Soft conflict.} A soft conflict occurs when demand policies enforced by different users for a specific device have overlapping device conditions.
    \item \textit{Restriction conflict.} Restriction conflict occurs when the restriction policy for a device is being disputed by the restricted user.
\end{enumerate}

\noindent\textbf{Device condition.} We consider \textit{device condition} as the set of rules assigned to a device by the user to perform a specific task in a SHS. For instance, if the user configures a smart light to switch on at sunset, the specified time is considered as the device condition. 

%% file: problem.tex
\section{Problem and Threat Model}\label{problem}
We introduce the challenges of an access control mechanism and articulate the threat model.

\begin{figure}[t!]
\centering
\includegraphics[width=0.7\columnwidth]{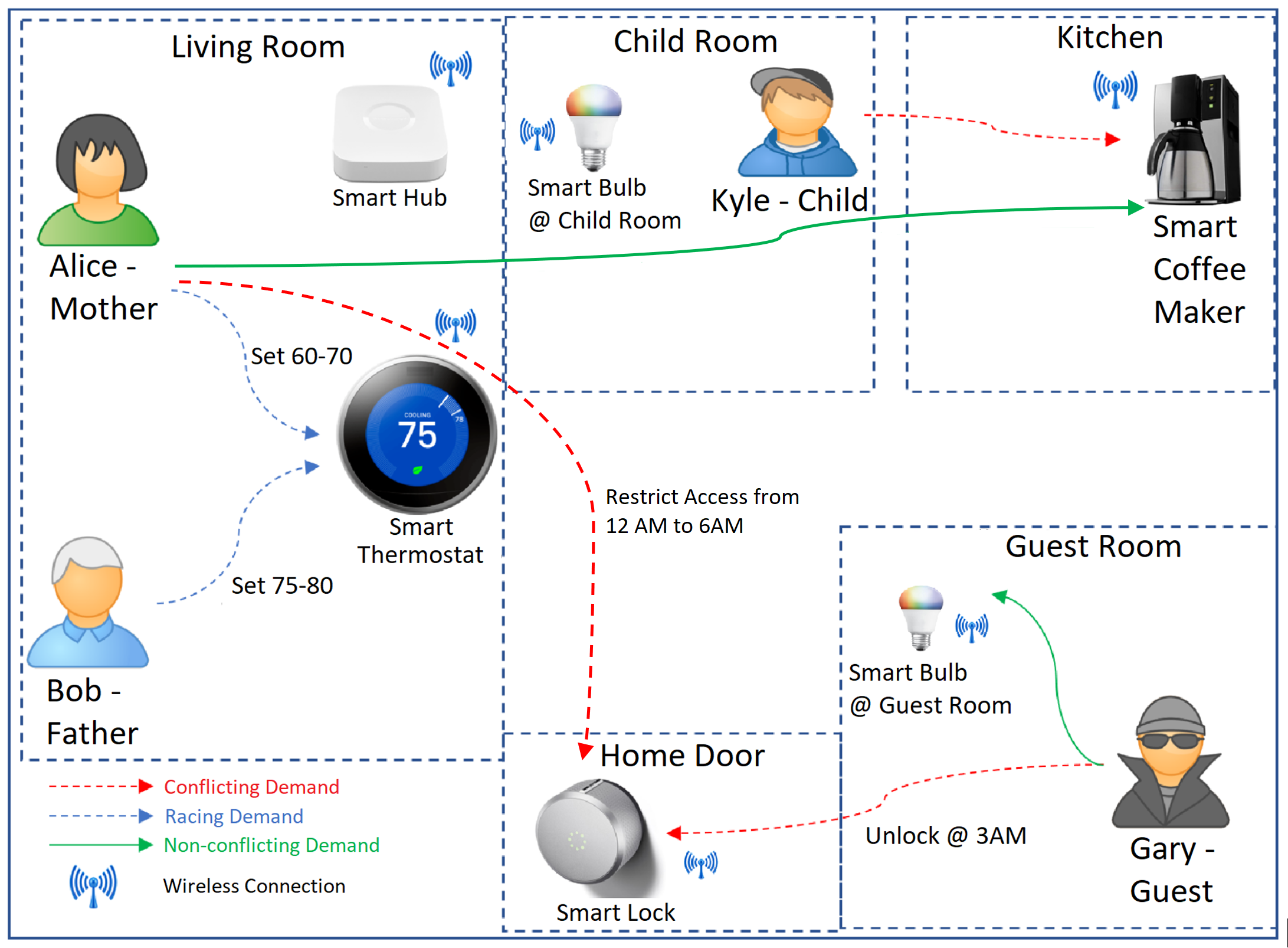}
\vspace{-0.1in}
\caption{Sample smart home with multiple users attempting to control multiple devices with conflicting demands. } 
\vspace{-0.2in}
\label{fig:smarthome}
\end{figure} 

\subsection{Problem Definition}
We assume a SHS (S) similar to the one depicted in the Figure~\ref{fig:smarthome}. In S, four different users – Bob (father), Alice (mother), Kyle (child), and Gary (guest) interact with the smart devices. 
Bob and Alice are the owners and all four users control the devices via the controller app. Here, the term \textit{access to the SHS
} refers to the ability of controlling the devices, configuring the system (add/delete devices), and adding new users. We assume the users perform the following conflicting activities- (1) Bob and Alice configure the smart thermostat to different overlapping values at the same time (soft conflict), (2) Alice wants to limit access of smart lock after midnight while Gary wants to have access (hard conflict), (3) Alice wants Kyle to have access to the smart light only while Kyle is present and restrict other devices (restriction conflict and location access). To address these, we propose \projectname, a fine-grained access control system for the smart home that allows users to resolve the conflicting access control demands automatically, add new users, select specific devices to share, limit the access to specific users, and prevent undesired user access based on user location in the system.

\subsection{Threat Model}
\projectname considers undesired access control decisions that may arise from existing coarse-grain solutions. For instance, a new user automatically gets full access to the system (i.e., over privileged control) which may lead to \textit{undesired device access}. Also, \projectname considers legitimate smart home users \textit{trying to change the system settings without authorization} (e.g., overriding existing system by installing new apps) that may result in undesired device actions such as installing unknown apps and overriding device conditions (i.e., privilege abuse), even deleting device owners from the system (i.e., privilege escalation). Furthermore, \projectname considers threats that arise from \textit{inadequate}, \textit{inaccurate}, or \textit{careless access control} to multi-user multi-device smart home (i.e., transitive privilege). In fact, access to a SHS granted to unknown parties by an authorized user other than the owner may escalate to additional threats (i.e., unauthorized device access), that \projectname also considers as malicious activity. Also, if a temporary guest is not timely removed from the system by the authorized user, it may lead to malicious activities such as sensitive information leakage.

We do not consider any unauthorized user access due to malicious apps installed in the system. We also assume that the SHS is not compromised, which means no malicious user is added automatically at the time of system installation as they are different problems from the contributions of \projectname.

%% file: technical.tex
\section{\textsc{KRATOS} Design} \label{sec:technical}

In this section, we present the architecture of the \projectname and its main components. \projectname is a comprehensive access control system for multi-user SHS where users can express their conflicting demands, desires, and restrictions through policies. \projectname allows an authorized user to add new users and enforce different policies to smart devices based on the needs of users and the environment. \projectname considers all the enforced policies from authorized users and includes a policy negotiation algorithm to optimize and solve conflicts among users. In designing the \projectname framework, we consider the following design features and goals. 

\vspace{4pt} \noindent \textbf{User-friendly Interface.} An access control system should have a user-friendly interface to add or remove users and assign policies in the SHS. We integrate \projectname into the mobile app provided by smart home vendors to provide a single user interface to manage users and assign policies in the connected devices.

\vspace{1pt} \noindent\textbf{Diverse User Roles/Complex Relations.} In a SHS, users have different roles that an access control system needs to define. For example, a user having a parent role should be able to express controls on a user with a child role, while adults in the same priority class should be able to negotiate the access control rules automatically. To address this design feature, \projectname introduces user priority in the system to define user roles.

\vspace{1pt} \noindent \textbf{Conflict Resolution.} As discussed earlier, diverse needs in device usage result in conflicts among users in a shared SHS. The main challenge of an access control system in a SHS is to resolve these conflicts in a justified way. In addition, users in a multi-user SHS should agree with the outcome of conflict resolution provided by the access control system. \projectname uses a novel policy negotiation system to automatically optimize and resolve the conflicting demands among users and institute a generalized usage policy reflecting the needs of all the users. 

\vspace{1pt} \noindent \textbf{Policy Expiration.} In a SHS, temporary access for different smart devices might be needed for guests or occasional users. To automate the temporary access, \projectname offers policy expiration time for specific users. \projectname automatically deletes the devices access and the user from the system after the expiration period to avoid undesired access in the system.

\vspace{1pt} \noindent \textbf{Location-based Access.} SHSs offer remote access and users can control the devices within the smart home. \projectname uses the location of the users to trigger a device policy to limit device usage and meet diverse needs of the users. For example, the parents may want to restrict remote access of the kids for specific devices. \projectname only allows users to access location-restricted devices if he/she is connected to the home network. 

\vspace{1pt} \noindent \textbf{Expressive Control.} In a SHS, a user should be able to express the desired device settings easily. An access control system should provide a simple method to the users to express their diverse needs. \projectname introduces a unified policy language that covers different control parameters (e.g., role, environmental, time, device, location-specific expressions) in a SHS to understand the users' needs and control the devices accordingly. 

\vspace{1pt} \noindent \textbf{Unified Policy Enforcement.} All user commands \cite{saint} to the devices should go through an \textit{access control enforcement} layer to provide fine-grained access control in SHS. \projectname uses an execution module that checks all enforced policies before executing a user command in the SHS.

\begin{figure}[t!]
	\centering{\includegraphics[width=0.5\textwidth]{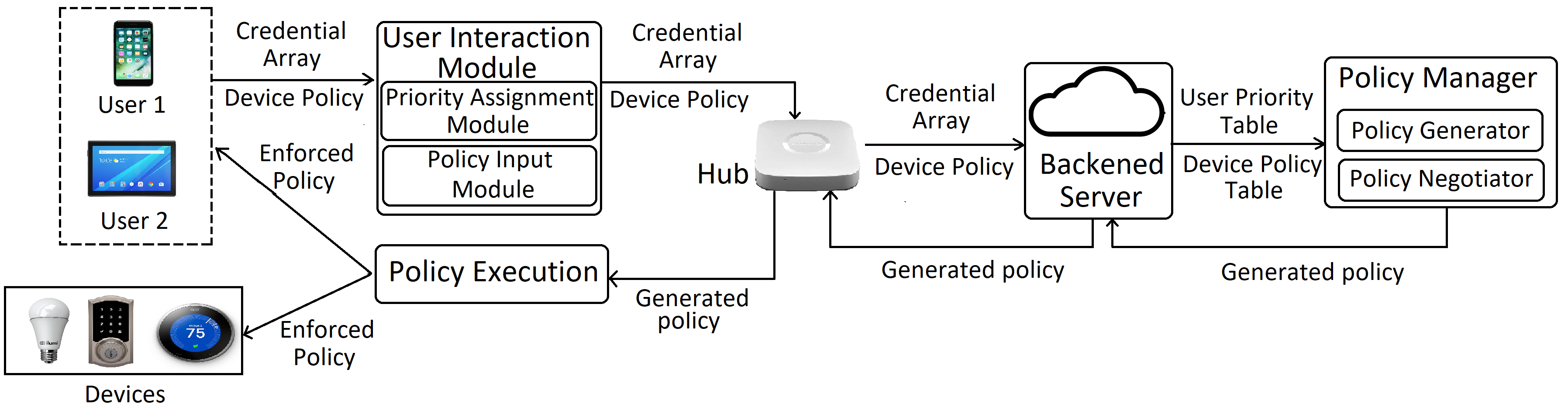}} 
	\vspace{-0.25in}
	\caption{Architecture of \projectname system.}
	\label{arc}
	\vspace{-0.25in}
\end{figure}

Figure~\ref{arc} shows the architecture of the \projectname system.  \projectname includes three main modules: (1) user interaction module, (2) back-end module, and (3) policy manager. First, the \textit{user interaction module} provides a user interface to add new users and assign priorities based on the user's role. This module also collects user-defined device policies for smart home devices. These device policies and priority assignment data are forwarded to the \textit{back-end module} via the smart home hub. back-end module captures these data and creates user priority and device policy list for the users. Lastly, the \textit{policy manager} module gathers user priorities and device policies from the generated lists and triggers policy generation and negotiation process. The following subsections details each module in \projectname and explains how policy generation and negotiation processes are initiated by \projectname.

\subsection{User Interaction Module}\label{rule}

The user interaction module collects priority assignment data and device policies from the users. It includes two sub-modules: priority assignment and policy input .

\vspace{1pt}\noindent\textbf{Priority Assignment Module.} The priority assignment module operates as a user interface to add new users and assign priorities. \projectname introduces a formal format to specify new users, illustrated as follows: $U_a = [A_{id},\ N_{id},\ P,\ D, \ T]$, where, $A_{id}$ is the unique ID of the commanding user, $N_{id}$ is the new user ID that is added in the system, $P$ is the priority level of the new user, $D$ is the permission to add or remove devices from the system, and $T$ is the validity time of the new user in the system. The user priority level is used in the policy generation module to resolve conflicting demands. For adding a new user and assigning priorities, we consider the following rules to avoid conflicts in the priority assignment.

\noindent\textbullet\hspace{1mm} Each user has an authority to add new users and assign a priority.

\noindent\textbullet\hspace{1mm} The Owner of the SHS will have the highest priority in the system.

\noindent\textbullet\hspace{1mm} Priority in the system is depicted with a numerical value. The lower the priority of a user, the higher is the level of priority. For example, the owner of the hub has the priority of ``0''.

\noindent\textbullet\hspace{1mm} Each user can only assign the same or higher value of the priority to a new user, e.g., a user with a priority of ``1'' can only assign priority of ``1'' or higher to a new user.

\noindent\textbullet\hspace{1mm} If two existing users add the same new user with a different priority level, the user with a higher priority level gets the privilege to add the new user.

\noindent\textbullet\hspace{1mm} If two existing users with the same priority level assigned different priority levels to a new user, the system notifies the existing users to fix a priority level of the new user.

\noindent\textbullet\hspace{1mm} Each user can assign permissions for adding or removing devices to a new user if the commanding user has the same permission. 

The priority assignment of \projectname can also be configured to define the roles of the users. For example, the SHS in Figure~\ref{fig:smarthome}, Alice and Bob (parents) can be assigned to priority 0, Gary (guest) can be assigned to priority 2, and Kyle (child) can be assigned to priority 3. We use this priority list to explain the functions of \projectname throughout the paper. In \projectname, administrator or homeowner obtains the privilege to define the priority-role mappings in the system. \projectname also allows the users to add temporary users by specifying validity time ($T$) of a user in the system. After the specified validity time, \projectname removes the user from the system automatically preventing any unauthorized access from a temporary guest. 

\begin{figure}[t]
\centering
\centering
\fontsize{8}{8}\selectfont
\resizebox{0.45\textwidth}{!}
{
{%
\begin{tabular}{p{1.9cm} p{8cm}}
@\textrm{U}$_1$: Alice &  \\
\textrm{U}$_2$\textendash & restrict smart thermostat between 60-70 degrees.\\  
\textrm{U}$_3$\textendash & restrict access to smart coffee maker.\\ 
\textrm{U}$_3$\textendash & allow access to smart bulb at the child room if \textrm{U}$_4$ is home.\\
\textrm{U}$_4$\textendash & allow access to smart bulb at the guest room.\\ 
\textrm{U}$_4$\textendash & restrict access to smart lock at the home door from 12:00 AM to 6:00 AM.\\
@\textrm{U$_2$}: Bob &  \\ 
\textrm{U}$_1$\textendash & restrict smart thermostat between 75-80 degrees.\\ 
\textrm{U}$_3$\textendash & allow access to smart bulb in the child's room between 7:00 PM and 7:00 AM \\ 
\textrm{U}$_4$\textendash & allow access to smart lock at the guest room and the home door \\  
@\textrm{U$_3$}: Kyle &  \\  
\textrm{U}$_1$\textendash & desires to access smart bulb at the child room\\  
\textrm{U}$_1$\textendash & desires to access coffee maker\\  
@\textrm{U$_4$}: Gary &  \\  
\textrm{U}$_1$\textendash & desires to access smart lock at the guest room and the home door at 3 AM\\  
\textrm{U}$_1$\textendash & desires to access the smart bulb in guest room\\  
\end{tabular}%
}
}
\vspace{-0.1in}
\caption{An example demand and restriction requirements of users in Figure~\ref{fig:smarthome}.}
\vspace{-0.2in}
\label{fig:example-control}
\end{figure}

\vspace{1pt}\noindent\textbf{Policy Input Module and Access Policy Language.} Policy input module provides an interface to the users for assigning policies in smart home devices. All the authorized users can choose any installed device and create a device policy using this module. To define the device policies, \projectname introduces a formal access control policy language for the SHS to express complex user preferences (e.g., users' demands, desires, and restrictions) by utilizing an existing open-source smart home ecosystem (e.g., Samsung SmartThings). Each user defines a policy about their preferences for each smart device and any restriction over others' accesses in the SHS. For instance, sample policies for the smart home of four users shown in Figure~\ref{fig:smarthome}, where each user defines her requirements for other users in a smart home with the thermostat, bulbs, lock, and coffee maker, are shown in Figure~\ref{fig:example-control}. The criteria defined by the users are used throughout this sub-section to construct their policies.

\vspace{1pt}\noindent\textbf{Policy Structure.} \projectname represents the policies as collections of clauses. The clauses allow each user to declare an independent policy for their demands and other users. The clauses have the following structure:   $ \langle \textmd{users}\rangle :\langle \textmd{devices}\rangle : \langle \textmd{conditions}\rangle : \langle \textmd {actions}\rangle  $. The first part of the policy is \textit{users}, which includes the information of both policy assigner and assignee. The second part, \textit{devices}, specifies the device or a list of devices included in this statement. \projectname uses device ID assigned by the SHS to distinguish device-specific policies in a multi-device environment. The third part, \textit{conditions}, is a list of device conditions defining different control parameters (time-based operation, values, etc.) based on the capabilities of the smart devices. For instance, a user may define a condition where only a pre-defined range of commands or only a certain time-window is matched. The final part of the policy is  $\langle \textmd {action}\rangle$ which states the clause type, \textit{demand}, \textit{restrict}, or \textit{location}. We note that the \projectname's policy language allows users to define multiple clauses. For instance, a user may restrict a distinct subset of smart home devices for different conditions and different users. A sample policy scenario is illustrated in Figure~\ref{fig:example-policy}. 

\begin{figure}
\centering
{\footnotesize{

\begin{align*} 
  @\textsf{U$_1$} ~~ & \textsf{restrict ::  ~ : thermostat$_1$ : temperature $\notin [60-70]$ ;}  \\
     &\textsf{restrict ::  U$_3$ : coffeemaker : ~ ;}  \\
     &\textsf{location ::  U$_3$ : bulb$_3$ : location $\in [Home]$;}  \\
  &\textsf{demand ::  U$_4$ : bulb$_{4}$ : ~ ;}  \\
  &\textsf{restrict ::  U$_4$ : lock$_1$ : time $\notin [6:00am-9:00pm]$;}  \\
  @\textsf{U$_2$} ~~ & \textsf{restrict ::  ~ : thermostat$_1$ : temperature $\notin [75-80]$ ;}  \\
  &\textsf{demand ::  U$_3$ : bulb$_3$ : time $\in [7:00pm-7:00am]$;}  \\
  &\textsf{demand ::  U$_4$ : lock$_1$, lock$_4$ : ~;}  \\
  ... &~~~~~~~~~~~~~~~~~~~~~~~~~~~~~ ...
  \end{align*} 
}}
 \vspace{-0.3in} 
\caption{Sample policy clauses to partially implement demands and restrictions shown in Figure~\ref{fig:example-control}. } 
\vspace{-0.4in} 
 \label{fig:example-policy} 
 \end{figure}

\subsection{Back-end Module} 
The user interaction module collects the user credentials and device policies generated using the access policy language. It then forwards them to the back-end module where these data are stored and formatted for policy generation and negotiation. The back-end module has two functionalities: (1) generating user priority list, and (2) generating device policy list.

\vspace{1pt}\noindent\textbf{User Priority List.} The back-end module collects the credential arrays and creates a database for authorized users and their assigned priorities. All the credential arrays are checked with the priority assignment rules (explained in Section~\ref{rule}) and sorted as valid and invalid priority assignments. For each invalid priority assignment, the back-end module notifies the users who initiated the priority assignment. The back-end module also checks the validity of the users added in the user priority list based on the specified time in the credential arrays. The back-end module automatically removes user with expired validity and updates the user priority table. 

\vspace{1pt}\noindent\textbf{Device Policy List.} The back-end module accumulates all the policies assigned by the users and creates a database based on the device ID. As explained in Section~\ref{rule}, the access policy language assigns a device ID to determine the intended policy for each device. This list is updated each time a user generates a new policy.
\vspace{-4pt} 
\subsection{Policy Manager Module} 
The policy manager module collects the user priority list and device policy list from the back-end module and compares different user policies. This module consists of two sub-modules (policy negotiation module and policy generation module) to initiate the policy negotiation and generation processes. 

\vspace{1pt}\noindent\textbf{Policy Negotiation Module and Negotiation Algorithm.} The policy negotiation module compares all the user-defined policies and detects different types of conflicts based on user priorities and demands. Similar to traditional RBAC, \projectname uses assigned user roles and priorities  to understand the user needs in a smart home hierarchy. However, a smart home environment needs a more fine-grained approach than RBAC to address the conflicting scenarios based on users’ relationship, social norms, and personal preferences. To address these diverse needs, \projectname uses an automatic policy negotiation module to resolve conflicts in multi-user smart home environment. The policy negotiation module identifies types of conflicts based on user roles and priorities, categorizes the conflicts based on implemented policies, automatically decides whether a policy should be executed or not, starts a negotiation method between conflicting users using notification methods, and chooses an optimum operating point for both users upon mutual agreement. For policy negotiation, \projectname considers two essential research questions: (1) How does \projectname handle the policy conflicts between users with the same and different priority levels?, and (2) How does \projectname handle restriction policies without affecting smart home operations?, In the following, we address these questions.

The policy negotiation algorithm of \projectname processes all the policies and computes the negotiated results by modeling the users' authorities (classes, roles) in a multi-layer list. User authorities are split into ordered classes. Class 0 has the highest priority, and a higher class number means a lower priority. Each class may include a list of users (or roles as roles are just a set of users). Users at the same priority class shares the same priority. \projectname considers three types of conflicts (hard, soft, and restriction conflicts) between user policies after users are classified into authorities.

When two different policies include clauses of the same user's access for the same device, there can be an \textit{interference} between those clauses. Any such possible interference is further checked to disclose the potential conflicts. In this, hard conflicts can happen when two interfering clauses dictate different actions for some overlapping cases or dictate the same action for never overlapping cased. In other words, when policies have no possible way of cooperation or compromising, \projectname detects a hard conflict. However, if the same action exists with some common overlap while opposite actions never occur together, such interference is a soft conflict. Moreover, hard and soft conflicts are further categorized as \textit{Priority Conflicts} or \textit{Competition Conflicts} based on the priority of policy owners. When the conflict happens between users' policies who have different priority classes, \projectname defines a \textit{hard or soft priority conflict}. However, if the users have the same priority, \textit{hard or soft competition conflicts} happens. For hard priority conflicts, \projectname enforces the policy defined by the user with higher priority. In hard competition conflicts, \projectname initiates a negotiation process between the users with an average operational condition calculated from both policies. If the users mutually agree with an average operational condition, \projectname creates a new policy for the targeted device. In case of no mutual operation condition, \projectname notifies the higher priority user/admin to resolve the dispute with a common policy. In the case of both soft priority and soft competition conflicts, the result of the negotiation process of \projectname is a new clause with common set of conditions. If any interference is caused by the nature of action requested in two different policies, \projectname detects a restriction conflict in the system. By incorporating these with hard, soft, and restriction conflicts, \projectname overall implements five distinct conflict types (details in Appendix~\ref{policyalgorithm} and~\ref{negotiation}).

\vspace{1pt}\noindent\textbf{Policy Generation Module.} The goal of the policy generation module is to construct valid policies that reflect the demands and restrictions of all authorized users based on the device policies generated in the user interaction module. The generated policies are passed to the back-end module and stored in a database. Thereafter, these policies are enforced in smart devices. The negotiated policies computed by the policy negotiation algorithm are converted into enforceable access control rules. The negotiated policy clause, $\Psi =  \{P,U,D,\mathcal{C},A\}$, has a 5-tuple format  and is indeed well suited for existing attribute-based access control (ABAC) systems.  Thus, \projectname uses an ABAC-like enforcement for the final generated rules. Here, the policy, $B$, is the set of \{action, subject, resource, constraints\} tuples for a negotiated device policy. An example of mapping a sample policy to ABAC rule through a transformation function can be illustrated as follows: 
\begin{equation}
\footnotesize
\begin{split}
ABAC(\Psi_i)\ \coloneqq \{B \mid \operatorname{action}\ (B) = \operatorname{A_i}
                           & \land \operatorname{subject}\ (B) = \operatorname{U_i}\\
                           &\land \operatorname{resources}\ (B) = \operatorname{D_i} \\
                           & \land \operatorname{constraints}(B) = \operatorname{T_{\mathcal{C}_i}}\}
\end{split}
\end{equation}
where $T_C\ \coloneqq\{c \mid$ c\ satisfies the same conditions of C in mapped attributes into ABAC policy. Here, $ABAC(\Psi_i)$ holds a direct translation of actions, subjects, resources, and constraints. We develop an ABAC-like rule generator that enforces the rules in a control device. The generator is integrated into the hub device as a unified enforcement point.

\vspace{-0.2cm}
\subsection{Policy Execution Module}
Policy execution module enforces the final policies generated from the policy negotiation process. Smart home devices can be controlled through a mobile phone controller app or by installing different device-specific apps in the system (e.g., Samsung SmartThings). Policy execution process appends the generated policies in the smartphone controller app or the installed smart home apps. To append the policies, \projectname adds conditional statements to the app source code to enforce the policies. When a user tries to change the state of the device, the app asks the policy execution module to check in the policy table generated by a policy generator. If an acceptable condition is matched, the policy execution engine returns the policy to the app and creates a binary decision (true for the accepted policy and false for the restricted policy) in the conditional branches. Based on the decision enforced by the policy execution engine, the user command in a smart home app is executed.

%% file: implementation.tex
\section{\textsc{KRATOS} Implementation} 
\label{sec:implement}

We implemented \projectname in Samsung SmartThings platform which has the largest market share in consumer IoT, supports highest number of off-the-shelf smart home devices, and open-source apps~\cite{smartthingbusiness}. 

\vspace{1pt}\noindent\textbf{Implementation and Data Collection.}
We setup a SHS to test the effectiveness of \projectname. We used Samsung SmartThings hub and connected multiple smart devices and sensors to the hub. The complete list of devices in our SHS is provided in Appendix~\ref{devicelst}. The setup included four different types of devices: smart light, smart lock, smart thermostat, and smart camera, which are some of the most common smart home devices used in SHSs~\cite{common}. We also used three different types of sensors: motion, temperature, and contact sensors to provide autonomous control. Further, we collected data from 43 different smart home users. We selected the participants by conducting an institution-wide open call for participation and flyers for community outreach. We obtained the necessary Institutional Review Board approval for collecting data from real-life smart home users. While selecting participants for our study, we considered several features: (1) owns more than one smart home devices, (2) shares smart home environment with multiple users (e.g., parents, partners, friends, or housemates), (3) diverse user roles (working adults, housewives, young adults, student, etc.), and (4) beginner level knowledge on using smart home devices. The participants were grouped into 14 different groups and asked to choose their roles in a SHS. First, we recorded different conflicting scenarios experienced by the users and asked them to use \projectname to assign device policies. We investigated several multi-user scenarios for the policy generation and negotiation processes as detailed below:

\begin{figure}[t!]
\vspace{-0.1cm}
\centering
\subfloat[User Management]{\includegraphics[width=0.17\textwidth]{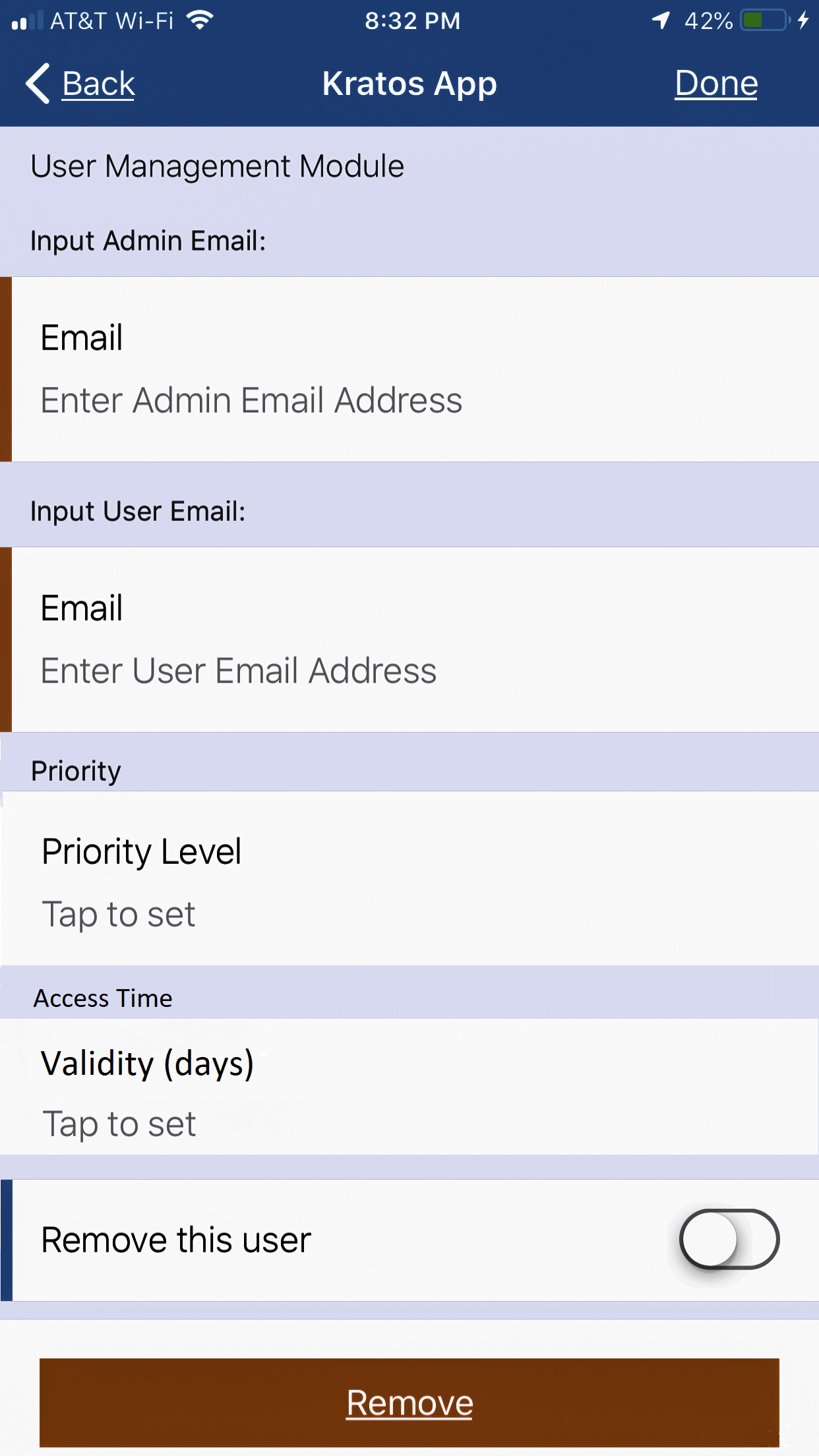}\label{fig:f10}}
\hspace{0.3in}
\subfloat[Policy Management]{\includegraphics[width=0.17\textwidth]{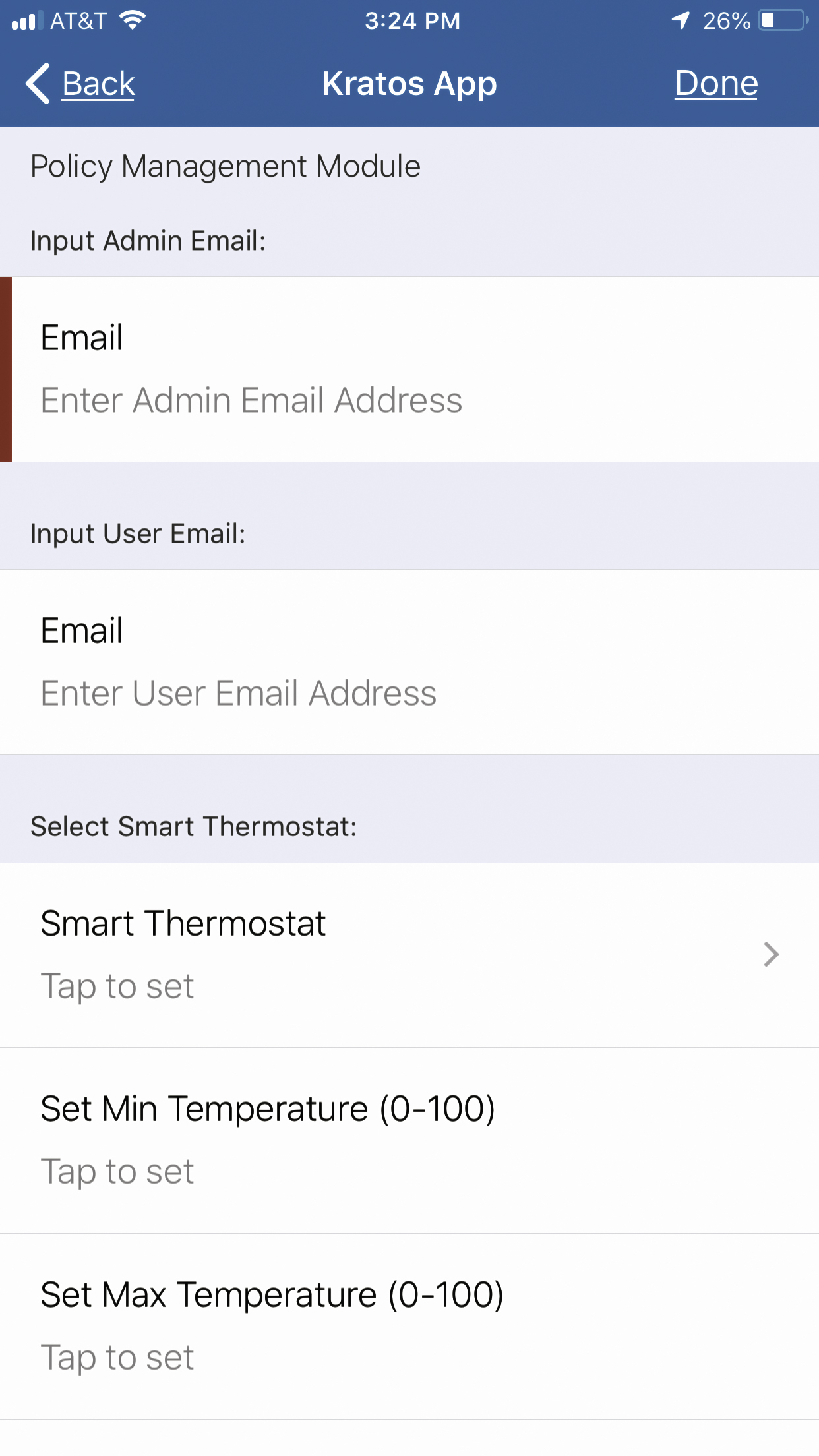}\label{fig:f11}}
\vspace{-0.1in}
\caption{User interfaces of \projectname.}
\label{kratosapp}
\vspace{-0.2in}
\end{figure}

\vspace{1pt}\noindent\textit{Scenario 1: Multiple policies for the same device.} We selected common devices (e.g., smart thermostat) and enforced different policies set by multiple users. Users assigned demand and restriction policies in the system for the same device. We collected 44 sets of policies (a set of policy includes at least two policies from multiple users) which included 13 hard, 17 soft, and 8 restriction conflicts. 

\vspace{1pt}\noindent\textit{Scenario 2: Multiple policies for different devices.} We used multiple devices from the same device category (e.g., smart light, smart lock, smart thermostat) to enforce different policies over the same type of devices. Here, we collected 48 sets of policies from 43 users which resulted in 15 hard, 22 soft, and five restriction conflicts.

\vspace{1pt}\noindent\textit{Scenario 3: Multiple apps for the same device.} In the SHS, we allowed users to install different apps to control the same device (e.g., smart light). For example, multiple users can configure a smart light with both motion and door sensors using different apps. We chose three different smart light apps available in SmartThings marketplace (light control with motion sensor, door sensor, and luminance level, respectively) and asked the users to install preferable apps and assign device policies accordingly. Here, we collected 35 sets of policies including 8 hard, 18 soft, and five restriction conflicts.

\vspace{1pt}\noindent\textit{Scenario 4: Single app for multiple devices.} We considered an individual app controlling multiple same types of devices in the SHS. We chose a single light controlling app to control four different lights and asked users to enforce device policies in different devices using one single app. We collected 32 sets of policies in this scenario which includes 12 hard, 15 soft, and 3 restriction conflicts. 

\vspace{1pt}\noindent\textit{Scenario 5: Temporary users in the system.} We considered a temporary user is added in the system and trying to access a smart light and smart lock after the access is expired for that specific user. We collected 30 sets of policies in this scenario.

\vspace{1pt}\noindent\textit{Scenario 6: Location-based access in the system.} In the location-based access control, we allowed multiple users to set location-based policies for a smart thermostat. Here, users are allowed to define both location-based restriction and demand policies. We collected 30 sets of policies in this scenario.

\vspace{1pt}\noindent\textbf{Malicious scenarios.} We also implemented five real-life threats in our SHS to generate malicious data and further evaluate the effectiveness of \projectname (more details in Section~\ref{sec:evaluation}) . For Threat-1 (Over privileged controls), we asked the users to add restriction clauses to the smart thermostat and asked the restricted users to change the temperature. For Threat-2 (Privilege abuse), we asked a newly added user with lower priority to install a new app in the smart home and trigger a smart camera. Threat-3 (Privilege escalation) is presented by a scenario where a new user changed the lock code of a smart lock and removed the smart lock from the environment. For Threat-4 (Unauthorized access), we added a temporary authorized user with limited priority and asked the users to control a smart thermostat outside their accepted time range. For Threat-5 (Transitive privilege), we asked the user with lower priority to add a new user with higher priority in the system.

\vspace{1pt}\noindent\textbf{User Interface.} We built a SmartThings app that represents the user interaction module described in Section~\ref{sec:technical}. This app has two modules: user management and policy management. The user management module allows users to add new users and assign priorities. We define five different roles and priority levels in \projectname (i.e., father/owner - priority 0, mother/owner - priority 0, adult - priority 1, guest - priority 2, child - priority 3). These roles and priorities can be assigned by the smart home owner or by authorized users with the same or higher priority to the one being assigned. Upon created a new role/priority, the information is sent and stored in the backend server. In the policy management module, users select devices and create new policies. \projectname provides options to add either general device policies (intended for all existing users) or policies that apply only to specific users. \projectname allows users to use different device conditions (operation-based, time-based, value-based, etc.) to define the policies. As our implementation environment had devices that only allows time-based and value-based conditions, we classified the policies in three different possible categories: (1) time-based device policy, (2) value-based device policy, and (3) time-value-based policy. The policies for different devices in our implementation can be represented by a device policy array: Device Policy, P = $\{U,\ D,\ C_1,\ C_2,\ R\}$.

\noindent\textbullet\hspace{1mm} \textit{User ID (U):} The first element of the policy array is to identify the policy assignee. We utilized the user email as a personal identifier in our implementation.

\noindent\textbullet\hspace{1mm} \textit{Device ID (D):} SmartThings assigns a unique device ID for each installed device which was used for the devices and policies.

\noindent\textbullet\hspace{1mm} \textit{Time conditions ($C_1$):} Users could assign a start time and an end time for any device action in the policies. For example, a smart light can be accessed from sunset to sunrise only. 

\noindent\textbullet\hspace{1mm} \textit{Value conditions ($C_2$):} Users could assign a maximum and minimum value to specify an acceptable range to control a device functionality. For example, a user can set the operational range of a thermostat from 68$^{\circ}$F  to 70$^{\circ}$F. 

\noindent\textbullet\hspace{1mm} \textit{Restricted User (R):} High-priority Users could define the restriction policy for a specific lower-priority user by adding the user ID to the restricted user's list. Users could also assign general policies (Section~\ref{defn}) for the devices by assigning '0' in this field.

Figure~\ref{kratosapp} shows the user interface of \projectname. We implemented \projectname as a customized smart app in Samsung SmartThings platform. We built the \projectname app in \textit{Groovy} platform and installed the app using SmartThings web interface. As Samsung allows each users to install customized apps in same smart environment using the web interface, \projectname app can be easily installed in each user's controller device in multi-user smart environment. Each user can use official SmartThings app in the controller device to use \projectname app to asign new users and device policies. The information of new users and device policies are forwarded to the policy generator via the backend server for generating final device policies. 

\vspace{1pt}\noindent\textbf{Policy Enforcement.} The final step during implementation is to enforce the generated policies by \projectname. We utilized 10 different official SmartThings apps that control 17 different devices and installed them in the SHS. We installed all the apps and observed the user-specific policies generated in the policy generation module. We modified these apps to connect with the backend server and capture the generated policies from the policy generator. These policies were appended to the conditional statements inside the app to execute the policies. A sample modified app is given in Appendix~\ref{patch} to illustrate the steps to enforce policies in a SmartThings app.

%% file: evaluation.tex
\section{Performance Evaluation} 
\label{sec:evaluation}
We evaluate \projectname by focusing on the following research questions:
\begin{itemize}\setlength\itemsep{0.0em}
    \item[\textbf{RQ1}] How effective is \projectname in enforcing access control in multi-user scenarios while handling different threat models? (Sec~\ref{sec:usage-scenario})
    \item[\textbf{RQ2}] What is the overhead introduced by \projectname on the normal operations of the SHS? (Sec.~\ref{sec:performance})
\end{itemize}
\vspace{-0.2cm}
\subsection{Effectiveness}
\label{sec:usage-scenario}

In this sub-section, we present the experimental results of \projectname while enforcing access control in different multi-user smart home scenarios and threat models. We first considered a use case scenario to explain the results of \projectname in different smart home operations. Then, we considered six different utilization scenarios (explained in Section~\ref{sec:implement}) to evaluate the effectiveness of \projectname. 
\begin{table}[t!]
\renewcommand{\arraystretch}{0.84}
\centering
\fontsize{8}{8}\selectfont
\resizebox{0.5\textwidth}{!}{
\begin{tabular}{|p{1.6cm}|p{5cm}|p{4.5cm}|}
\hline
\textbf{\begin{tabular}[c]{@{}c@{}}Conflict type\end{tabular}}  & \textbf{\begin{tabular}[c]{@{}c@{}}Policy example\end{tabular}} &  \textbf{\projectname outcome}\\ \hline\hline
\multirow{4}{*}{\begin{tabular}[p]{@{}c@{}}Hard priority \\conflict\end{tabular}} & Alice (priority-1) and Bob (priority-2) set up the temperature range 60-70 and 75-80 respectively in the smart thermostat. & As Alice has higher priority, \projectname sets the thermostat to 60-70 and notifies the users with the decision\\ \hline
\multirow{7}{*}{\begin{tabular}[p]{@{}c@{}}Soft priority \\conflict\end{tabular}} &  Alice (priority-1) and Bob (priority-2) set up the temperature range 60-70 and 65-75 respectively in the smart thermostat. & \textbullet \hspace{1mm} As Alice has the higher priority, \projectname sets the thermostat to 60-70 and notifies Alice with common range (65-70). \par \textbullet \hspace{1mm} If Alice agrees with common range, \projectname sets the temperature range 65-70.\\ \hline
\multirow{7}{*}{\begin{tabular}[p]{@{}c@{}}Hard \\competition \\conflict\end{tabular}} & Alice (priority-2) and Bob (priority-2) set up the temperature range 60-70 and 75-80 respectively in the smart thermostat. & \textbullet \hspace{1mm} \projectname starts the negotiation with average range (67-75) and upon mutual agreement from the users set the range. \par \textbullet \hspace{1mm} If the users fail to agree, \projectname notifies higher level user/admin to decide the policies.\\ \hline
\multirow{4}{*}{\begin{tabular}[p]{@{}c@{}}Soft \\competition \\conflict\end{tabular}} & Alice (priority-2) and Bob (priority-2) set up the temperature range 60-70 and 65-75 respectively in the smart thermostat. & \projectname sets the temperature range 65-70 and notifies the users with updated policy.\\ \hline
\multirow{5}{*}{\begin{tabular}[p]{@{}c@{}}Restriction \\conflict\end{tabular}} & Alice (priority-1) set the temperature range 60-70 and restrict Bob (priority-2) to change the thermostat. Bob sets the temperature range 75-80. & \projectname sets the temperature range 60-70 and notifies Bob regarding restriction.\\ \hline
\multirow{6}{*}{\begin{tabular}[p]{@{}c@{}}Temporary \\ access\end{tabular}} & Alice (priority-1) added Gary (priority- 4) as a temporary user for 2 days. After 2 days, Gary tries to unlock the smart lock. & \projectname automatically detects the expired validity for smart home access and deletes Gary from authorized user list to prevent any undesired access.\\ \hline
\multirow{6}{*}{\begin{tabular}[p]{@{}c@{}}Location-\\based \\ access\end{tabular}} & Alice (priority-1) set up the temperature range 70-72 and restrict Kyle (priority-3) from using the smart thermostat remotely. Kyle sets the temperature range 74-76. & \textbullet \hspace{1mm} If Kyle is not in the home network, \projectname disregard Kyle's access policy. \par \textbullet \hspace{1mm} \projectname checks the location of both Kyle and Alice. If only Kyle is home, \projectname sets the temperature range 74-76. If both Kyle and Alice are home, \projectname sets the temperature range 70-72. \\ \hline
\end{tabular}
}
\caption{Different usage scenarios and outcomes of \projectname.}
\label{outcome}
\vspace{-0.4in}
\end{table}

To understand the performance of \projectname, we assume two users Alice and Bob using the same smart thermostat and assigning different policies according to their needs. This usage scenario may lead to conflicts in which case \projectname uses policy negotiation module to solve the conflicts. For instance, let us assume Alice and Bob has the same priority level which is 2 and assign temperature range 60-70 and 75-80 respectively. \projectname considers this as a hard competition conflict and starts the negotiation process with average range 67-75. If Alice and Bob both agree with the range, \projectname generates a new policy for the thermostat with the temperature range 67-75 and enforces this in the device. On the other hand, if Alice and Bob cannot agree, \projectname notifies a higher level user/admin to resolve this conflict by assigning a new policy for the device. We also consider a temporary user scenario in evaluating \projectname where Alice (priority-1) adds a temporary user Gary (priority-4) in the system for 2 days. After the validity period (2 days), Gary tries to access the smart home devices. However, \projectname automatically detects any expired validity of the users in the system and restricts the temporary users to access the system. Table~\ref{outcome} summarizes the outcome of \projectname in different usage scenarios. Table~\ref{policy} also shows the summary of policy conflicts and negotiations between smart home users in different multi-user scenarios explained in Section~\ref{sec:implement}. In Scenario-1, \projectname successfully negotiated 44 sets of policies collected from 43 users and executed the generated policies in the SHS. Average policy generation time including the policy negotiation was 0.68 seconds. In Scenario-2, \projectname evaluated 48 sets of policies in total with an average policy generation time of 1.2 seconds. In Scenario-3 and 4, \projectname manages 35 and 32 sets of policies with an average generation time of 0.86 and 0.48 seconds respectively. In Scenario-5, \projectname successfully manages 20 sets of policies and automatically detects unauthorized access for expired temporary access. For location-based access in Scenario-6, \projectname successfully manages 30 sets of policies and provides location-based acess to multiple users. \projectname also successfully resolves all the conflicts generated in different scenarios. In summary, \projectname successfully resolved the policy conflicts and created optimized final policies that could be executed within different smart home apps.

\begin{table}[b!]
\vspace{-0.3cm}
\renewcommand{\arraystretch}{0.82}
\renewcommand{\arraystretch}{1.3}
\centering
\footnotesize
\fontsize{8}{10}\selectfont
\resizebox{0.48\textwidth}{!}{
\begin{tabular}{|c|c|c|c|c|c|c|c|}
\hline
\textbf{\begin{tabular}[c]{@{}c@{}}Usage\\ Scenario\end{tabular}}&  \textbf{\begin{tabular}[c]{@{}c@{}}No. of \\policies\end{tabular}} & \textbf{\begin{tabular}[c]{@{}c@{}}No. of hard\\conflicts\end{tabular}} &\textbf{\begin{tabular}[c]{@{}c@{}}No. of soft\\conflicts\end{tabular}}&\textbf{\begin{tabular}[c]{@{}c@{}} Restriction\\policies\end{tabular}}& \textbf{\begin{tabular}[c]{@{}c@{}} No conflicts\end{tabular}}& \textbf{\begin{tabular}[c]{@{}c@{}}Average \\time (s)\end{tabular}} & \textbf{\begin{tabular}[c]{@{}c@{}}Success \\rate (s)\end{tabular}}\\
\hline
\hline
{\begin{tabular}[c]{@{}c@{}}Scenario-1\end{tabular}} & 44 & 13 & 17 & 8 & 6 & 0.68 & 100\%\\\hline
{\begin{tabular}[c]{@{}c@{}}Scenario-2\end{tabular}} & 48 & 15 & 22 & 5 & 6 & 1.2 & 100\%\\\hline
{\begin{tabular}[c]{@{}c@{}}Scenario-3\end{tabular}} & 35 & 8 & 18 & 5 & 4 &0.86 & 100\%\\\hline
{\begin{tabular}[c]{@{}c@{}}Scenario-4\end{tabular}} & 32 & 12 & 15 & 3 & 2 &0.48 & 100\%\\\hline
{\begin{tabular}[c]{@{}c@{}}Scenario-5\end{tabular}} & 30 & - & 12 & 6 & 12 & 0.2 & 100\%\\\hline
{\begin{tabular}[c]{@{}c@{}}Scenario-6\end{tabular}} & 30 & 10 & 8 & 8 & 4 & 0.32 & 100\%\\\hline
\end{tabular}}
\caption{\projectname's performance in different scenarios.}
\vspace{-1cm}
\label{policy}
\end{table}

We also evaluated the effectiveness of \projectname in preventing different threats in the SHS. We considered five different threats presented in Section~\ref{sec:implement}. We collected data from fifty malicious occurrences in total to evaluate \projectname against these threats. Table~\ref{threat} summarizes the performance of \projectname in identifying different threats. In each of these scenarios, \projectname detected the policy violation with 100\% accuracy and effectively notified the smart homeowner/policy assigner via push notifications. For Threat-1, \projectname achieves the lowest average detection and notification time 0.25 and 0.4 seconds respectively. To identify Threat-2 and 3, \projectname takes 0.4 and 0.47 seconds on average with average notification time 0.6 seconds. For Threat-4 and 5, the average detection time is 0.35 and 0.28 seconds, respectively. In summary, \projectname can detect different threats with 100\% accuracy and notify users with minimum delay.

\begin{table}[h!]
\renewcommand{\arraystretch}{0.82}
\centering
\footnotesize
\fontsize{8}{10}\selectfont
\resizebox{0.47\textwidth}{!}{
\begin{tabular}{|c|c|c|c|c|}
\hline
\textbf{\begin{tabular}[c]{@{}c@{}}Threat \\ model\end{tabular}} &  \textbf{\begin{tabular}[c]{@{}c@{}}No. of \\occurances\end{tabular}} & \textbf{\begin{tabular}[c]{@{}c@{}}Success \\ rate\end{tabular}} &\textbf{\begin{tabular}[c]{@{}c@{}}Average Detection \\ time (s)\end{tabular}}&\textbf{\begin{tabular}[c]{@{}c@{}} Average Notification \\ time (s)\end{tabular}}\\
\hline
\hline
{\begin{tabular}[c]{@{}c@{}}Threat-1\end{tabular}} & 10 & 100\% & 0.25 & 0.4 \\\hline
{\begin{tabular}[c]{@{}c@{}}Threat-2\end{tabular}} & 10 & 100\% & 0.4 & 0.6 \\\hline
{\begin{tabular}[c]{@{}c@{}}Threat-3\end{tabular}} & 10 & 100\% & 0.47 & 0.6 \\\hline
{\begin{tabular}[c]{@{}c@{}}Threat-4\end{tabular}} & 10 & 100\% & 0.35 & 0.52 \\\hline
{\begin{tabular}[c]{@{}c@{}}Threat-5\end{tabular}} & 10 & 100\% & 0.28 & 0.45 \\\hline
\end{tabular}}
\caption{Performance of \projectname against different threats.}
\label{threat}
\vspace{-0.3in}
\end{table}

\vspace{-0.2cm}

\subsection{Performance Overhead}\label{sec:performance}
\begin{figure*}[h!]
\vspace{-0.2in}
\centering
    \subfloat[]{\includegraphics[width=0.2\textwidth]{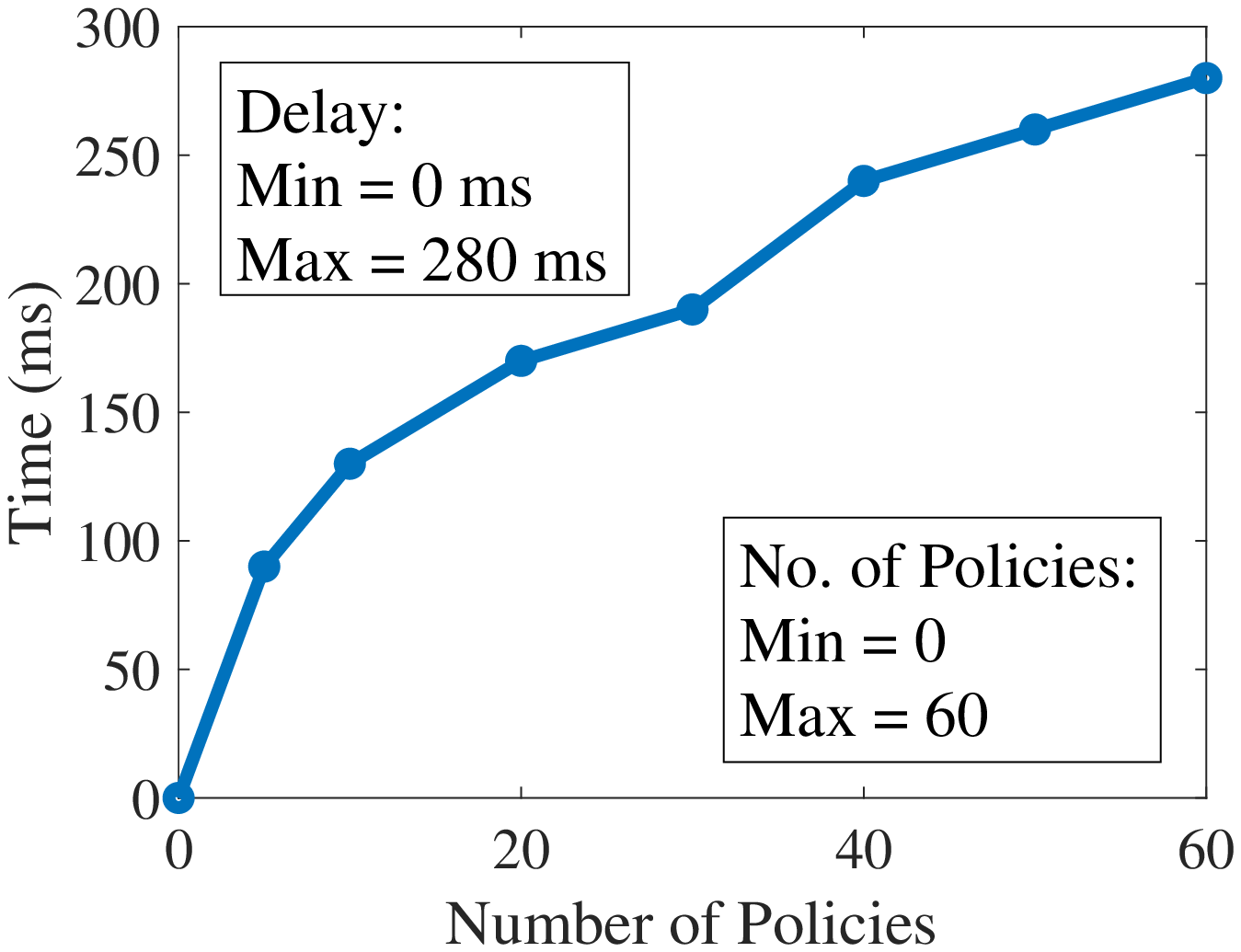}\label{fig:f3}}
    \subfloat[]{\includegraphics[width=0.2\textwidth]{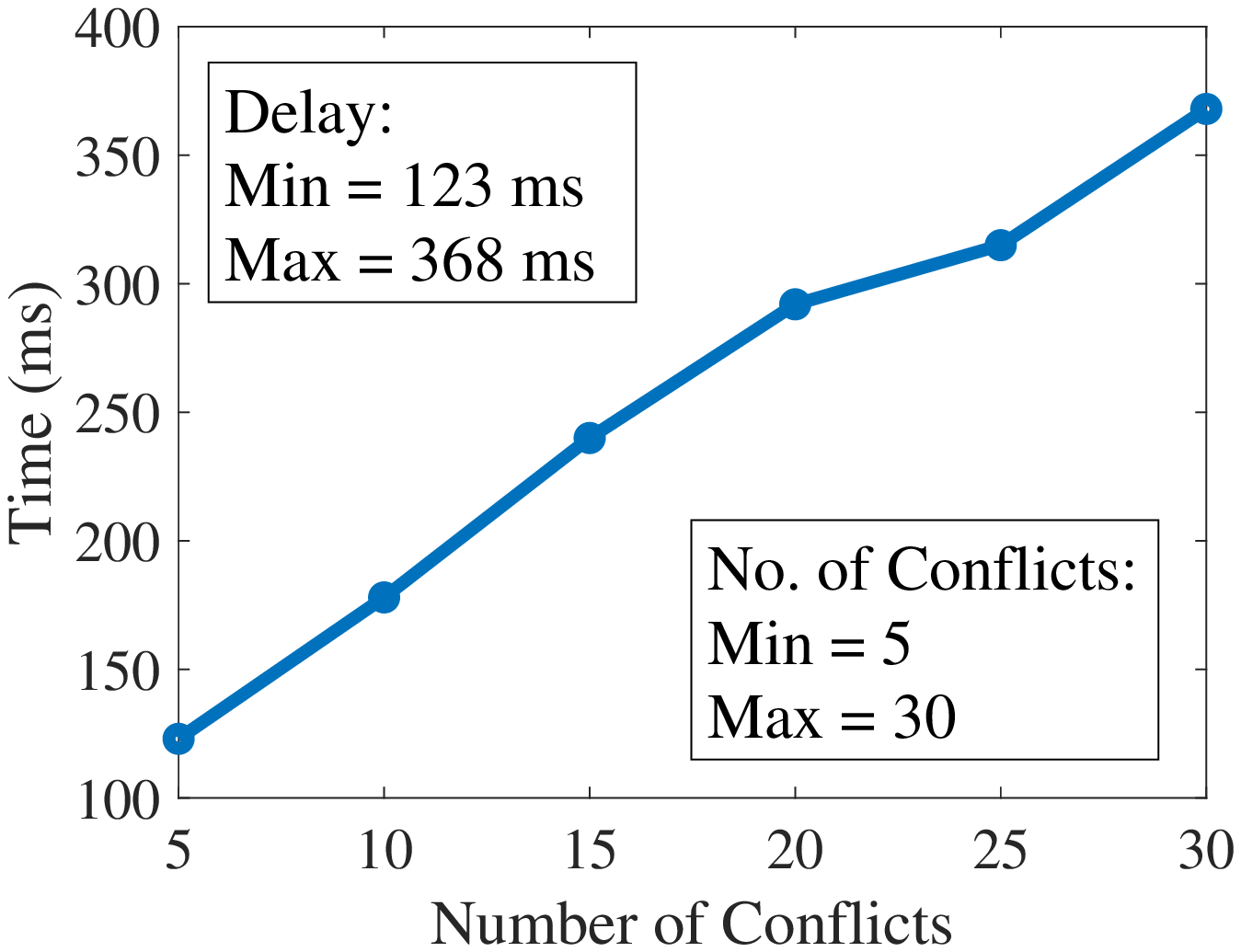}\label{fig:f4}}
    \subfloat[]{\includegraphics[width=0.2\textwidth]{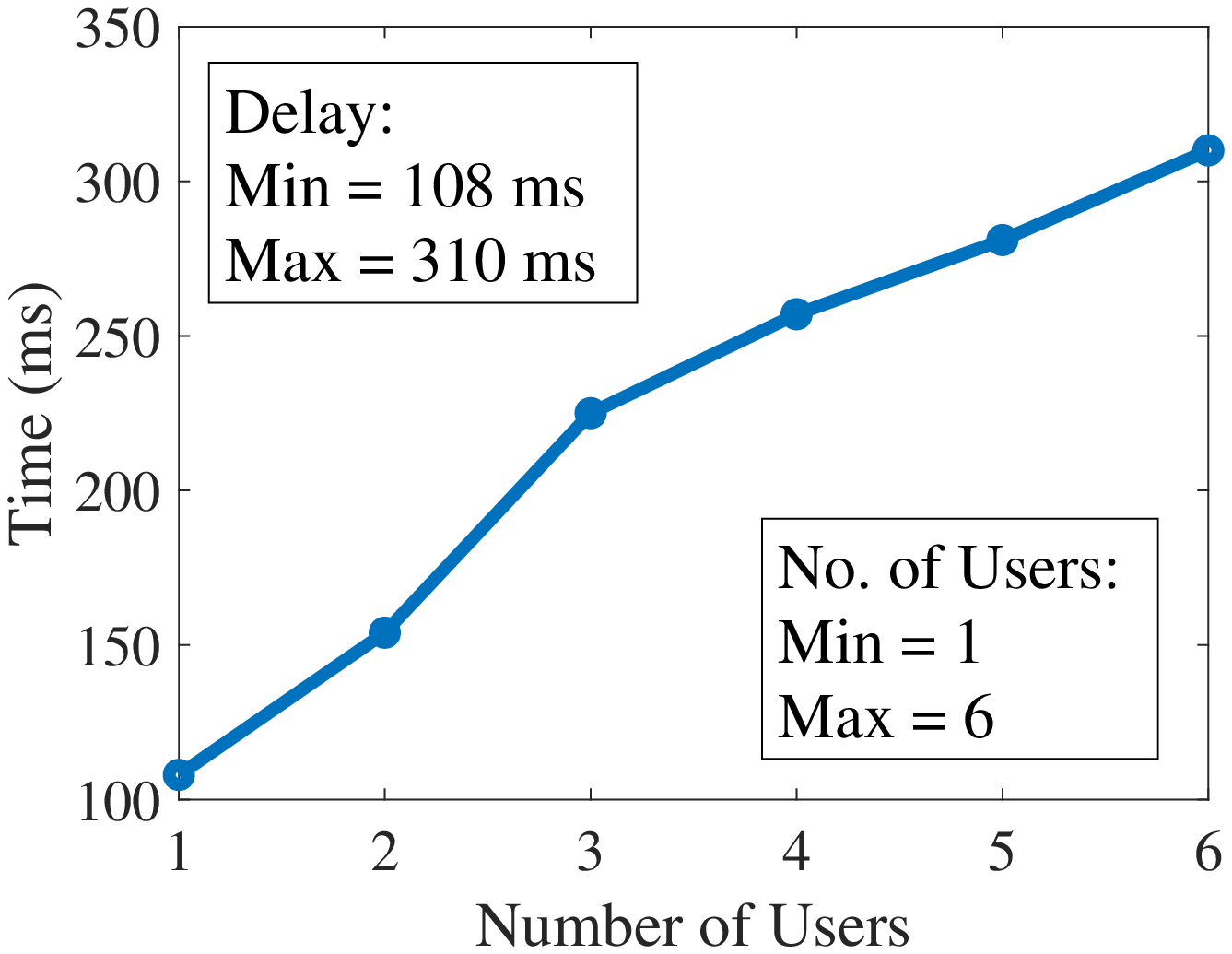}\label{fig:f5}}
    \subfloat[]{\includegraphics[width=0.2\textwidth]{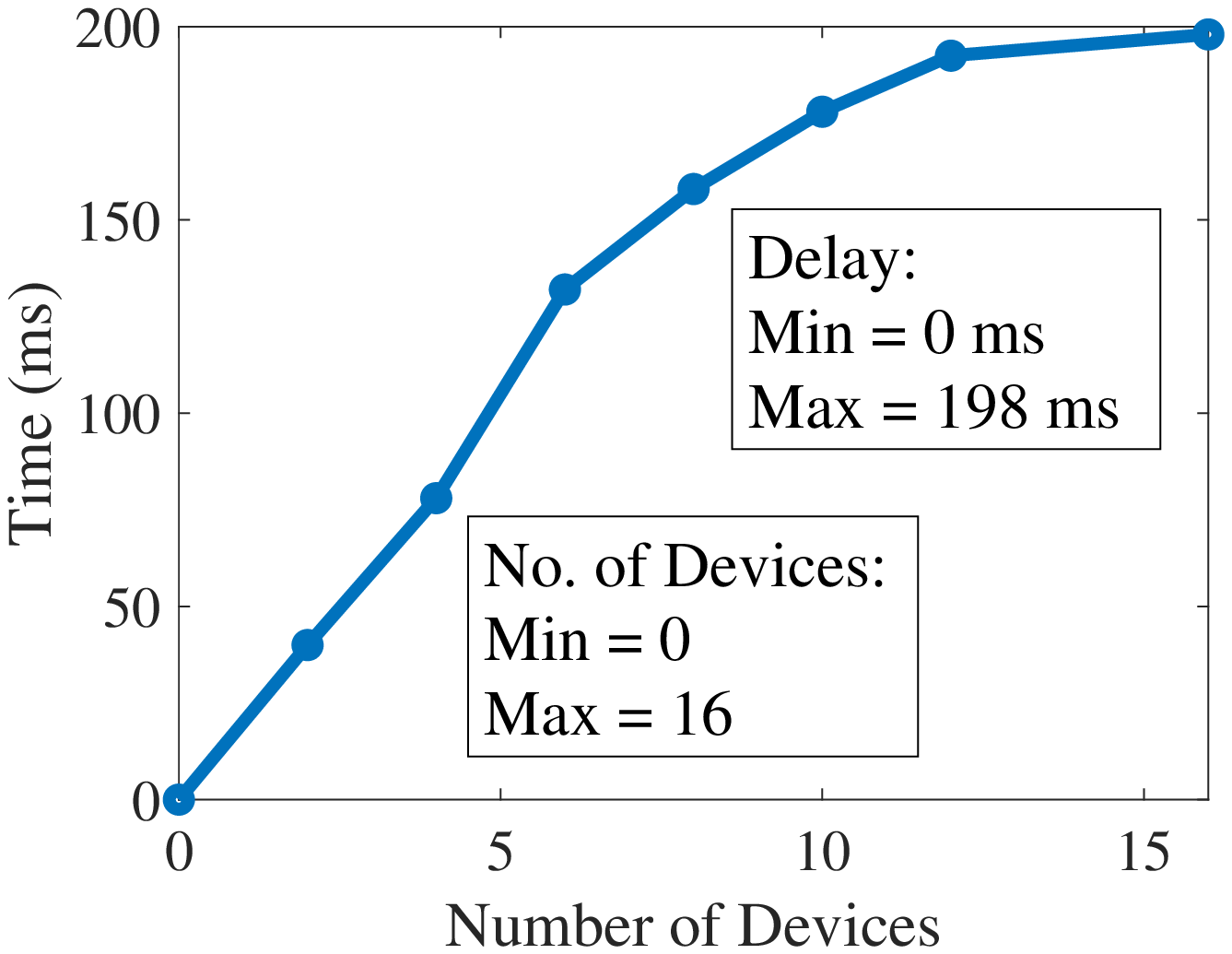}\label{fig:f6}}
    \vspace{-0.1in}
    \caption{Impact of different evaluation parameters on \projectname's performance: (a) number of policies, (b) number of conflicts, (c) number of users, and (d) number of devices.}
    \label{timedelay}
\vspace{-0.1in}
\end{figure*}

We considered the following research questions to measure the performance overhead of \projectname:
\begin{enumerate} \setlength\itemsep{0.0em}
    \item[\textbf{RQ3}] What is the impact of  \projectname in normal operations of the SHS? (Table~\ref{policynego}) 
    \item[\textbf{RQ4}] What is the impact of \projectname in executing a user command in the SHS via the smart home apps? (Table~\ref{policyexecution}) 
    \item[\textbf{RQ5}] How does the impact of \projectname change with different parameters in the SHS? (Figure~\ref{timedelay}) 
\end{enumerate}
For different multi-user scenarios, we considered four different scenarios as explained in Section~\ref{sec:implement}.

\vspace{1pt}\noindent \textbf{Latency Introduced by \projectname.}  \projectname considers three different types of conflicts (hard conflicts, soft conflicts, and restriction policy) during policy generation and negotiation based on user priorities and policy types. These policy generation and negotiation processes normally introduce latency in the normal operations of a SHS and the smart apps to analyze given policies and solving conflicts. Table~\ref{policynego} illustrates the delay introduced by \projectname while handling the policy conflicts and negotiations. We note that the average negotiation time increases with the number of policies for all types of policy conflicts. For hard conflicts, the average negotiation time is 0.403 seconds for ten policies, which increases to 1.21 seconds for 30 policies. Because the hard conflicts require all the conflicted users to interact with the system to resolve the conflicts, it takes more time than soft conflict and restriction policies. For soft conflicts, the average negotiation time is 0.27 seconds for ten policies which increases to 0.73 seconds for 30 policies. For the restriction policies, the latency is introduced only when a low-priority user tries to assign policies to high-priority users. In this case, average negotiation times vary from 0.102 seconds to 0.25 seconds from 10 to 30 policies.

\begin{table}[h!]
\vspace{-0.1cm}
\renewcommand{\arraystretch}{1.2}
\fontsize{6}{6}\selectfont
\centering
\begin{tabular}{|c|c|c|}
\hline
\textbf{\begin{tabular}[c]{@{}c@{}}Conflict types\end{tabular}} &  \textbf{\begin{tabular}[c]{@{}c@{}}No. of Policies\end{tabular}} & \textbf{\begin{tabular}[c]{@{}c@{}}Average negotiation time (s)\end{tabular}} \\ \hline
\hline
\multirow{3}{*}{\begin{tabular}[|c|]{@{}c@{}}Hard conflict\end{tabular}} & 10 & 0.403 \\\cline{2-3}
                             & 20 & 0.715 \\ [0ex]\cline{2-3}
                              & 30 & 1.21 \\[0ex]\cline{2-3}\hline
\multirow{3}{*}{\begin{tabular}[c]{@{}c@{}}Soft conflict\end{tabular}} & 10 & 0.27 \\\cline{2-3}
                             & 20 & 0.53 \\\cline{2-3}
                             & 30 & 0.73 \\\cline{2-3}\hline
\multirow{3}{*}{\begin{tabular}[c]{@{}c@{}}Restriction Policy\end{tabular}} & 10 & 0.102 \\\cline{2-3}
                             & 20 & 0.117\\\cline{2-3}
                             & 30 & 0.25 \\\cline{2-3}\hline
\end{tabular}
\caption{Overhead of \projectname in handling policy negotiations.}
\vspace{-0.3in}
\label{policynego}
\end{table}

\vspace{1pt}\noindent \textbf{Impact of \projectname on Executing User Commands.} As the policies in \projectname are enforced in the smart apps installed via the controller device (e.g., smartphone and smart tablet), it introduces overhead in the controller devices while installing the apps and executing users' command. Table~\ref{policyexecution} depicts the impact of \projectname on executing user commands based on generated policy. Here, we used eight different apps to measure the performance overhead of \projectname. We also considered three types of constraints on the policies: time constraint, value constraint, and both time and value constraints. Time constraint refers to the specific time range for the desired action of a smart device (e.g., turning on lights at sunset) while value constraint refers to the specific range of inputs to a smart device (e.g., the temperature of the smart thermostat). With no policy enforced on a device, the average time to install an app and execute user command is 1.3 seconds with 1.75\% and 1.6\% of CPU and RAM utilization, respectively. For time constraints and value constraints, the average time is 1.72 and 1.46 seconds, respectively. Average CPU and RAM utilization are almost similar for both time and value constraints (2.1-2.2\% and 2.25-2.6\%, respectively). For both time and value constraints, the average execution time increases to 1.92 seconds. The CPU and RAM utilization also increases to 2.5\% and 2.82\%, respectively. Considering the CPU and RAM available in modern smartphones and tablets, the overhead introduced by \projectname can be considered negligible~\cite{203854, sikder2019context, sikder2019patent}.

\begin{table}[h!]
\vspace{-0.2cm}
\renewcommand{\arraystretch}{0.95}
\centering
\fontsize{6}{8}\selectfont
\begin{tabular}{|c|c|c|c|}
\hline
\textbf{\begin{tabular}[c]{@{}c@{}}Type of policy\end{tabular}} &  \textbf{\begin{tabular}[c]{@{}c@{}}Avg. time (s)\end{tabular}} & \textbf{\begin{tabular}[c]{@{}c@{}}Avg. CPU usage\end{tabular}} & \textbf{\begin{tabular}[c]{@{}c@{}}Avg. RAM usage\end{tabular}}\\ \hline
\hline
{\begin{tabular}[c]{@{}c@{}}No policy\end{tabular}} & 1.3 & 1.75\%  & 1.6\%\\\hline
{\begin{tabular}[c]{@{}c@{}}Time constraint\end{tabular}} & 1.72 & 2.2\% & 2.6\%  \\\hline
{\begin{tabular}[c]{@{}c@{}}Value constraint\end{tabular}} & 1.46 & 2.1\% & 2.25\% \\\hline
{\begin{tabular}[c]{@{}c@{}}Time and Value constraint\end{tabular}} & 1.92 & 2.5\% & 2.82\% \\ \hline

\end{tabular}
\caption{Overhead of \projectname in policy executions.}
\vspace{-0.33in}
\label{policyexecution}
\end{table}
\vspace{1pt}\noindent \textbf{Impact of Different Parameters on Performance Overhead.} \projectname considers different parameters in SHSs to define and execute device policies reflecting diverse user demands. Here, we observed the performance overhead of \projectname by changing various parameters. As policy generation and negotiation are executed at the backend server, \projectname does not pose any performance overhead to computational parameters (CPU and RAM utilization). The only noticeable change is observed in delay imposed by \projectname in the normal operation of the SHS. In Figure~\ref{timedelay}, the delay introduced by \projectname is shown based on the number of policies, conflicts, users, and devices. One can notice from Figure~\ref{fig:f3}, the delay introduced by \projectname increases with the number of policies generated by the users. \projectname introduces 90 ms delay in the SHS for five policies to execute a user command which increases to 280 ms delay for 60 policies. The delay increases linearly with the number of conflicts and users in the system (Figure~\ref{fig:f4} and Figure~\ref{fig:f5}). The highest delay to execute a user command is 368 ms, which occurs when the system includes 30 different policy conflicts. \projectname also takes 310 ms to execute a command with six different users presents in the system. This delay is the result of the overhead introduced by notifying different users about executing the command. For the number of devices, the delay introduced by \projectname becomes steady after adding 12 different devices in the SHS~(Figure \ref{fig:f6}).

%% file: benefits.tex
\section{Benefits and Future Work} \label{sec:benefits}

\subsection{Benefits of \textsc{Kratos}}
Consider a user, Bob, who defines himself as a technology savvy and enthusiastic entrepreneur homeowner. Bos's house is set with devices such as smart lock, thermostat, and fire alarm. Bob's is the head of a family of three members, including his wife Alice, and his teenager son Matt. Finally, Bob offers high-quality vacation rentals to Airbnb users.

\vspace{1pt}\noindent \textbf{Efficient Conflict Resolution.} With several devices shared among all household members (including the Airbnb tenant), Bob feels that there is an immediate need for some control mechanism that defines how all the smart devices are being set up and managed among the different users. However, despite trying devices and smart apps from different platforms (e.g., Samsung SmartThings, Google Home, etc.), Bob cannot find a feasible and user-friendly solution that consider the needs of the different users (e.g., Bob and Alice's priority is to keep the thermostat temperature as high as possible while Matt's idea is to have cooler temperature). \projectname offers a fine-grained access control mechanism for the SHS that allows Bob to provide access control based on the users' needs, demands, and priorities.

\vspace{1pt}\noindent \textbf{Multi-users/Multi-devices.} As mentioned before, Bob's setup comprises several devices with different levels of usability based on their impact on the quality of life of users and their contribution to the general security of the household. Additionally, users may have different levels of access based on Bob's and the household's best interests. Based on these scenarios, Bob expects an smart home access control system capable of managing multi-user and multi-device environments. \projectname realizes and offers an access control system where the administrator (i.e., Bob) can assign priority levels to the different devices and users. This allows control mechanisms to consider the importance of the various devices, but also the needs of the users based on admin's pre-defined priorities.

\vspace{1pt}\noindent \textbf{Suitability for Complex User Demands.} Users' demands can be very complex at times. For instance, in addition to the demands and interests of Bob, Alice, and Matt, new access control policies can be generated in case Bob decides to give some control to his Airbnb tenant Ed. Adding new users and devices to an already configured SHS increase the complexity due to new conflicts between users and device policies. To solve these issues, \projectname can actively analyze and solve policy conflicts through negotiations in an optimized fashion based on the different user and device priorities.

\vspace{1pt}\noindent \textbf{Inherent Security.} Bob has certain rules to protect his smart home ecosystem. First, security-related devices (e.g., smart lock) have the highest priority. Second, he would like to have strict and unique control over these devices, so no other user can change their settings or expected behaviors. Finally, users with the lowest priority (e.g., Ed) should not be able to add new devices, change SHS settings, etc. Our framework was designed to provide inherent security based on the specific user's needs. Specifically, \projectname offers the means to provide complex control and demands through comprehensive policy negotiation and conflict resolution.

\subsection{Future Work}
\noindent \textbf{User and Usability Study.}  
To understand the users' access control needs in a smart home environment, we will conduct a detailed user study considering user characterization, device sage patterns, smart home configuration references, and user preferences Also, as \projectname is a user-centric access control solution, usability of \projectname should be tested in a real-life smart home environment. We will conduct a usability study among smart home users to test \projectname based on different parameters such as user interface, 
acceptability of use, availability, user friendliness, notification system, and effectiveness. We will also develop tutorials and detailed user guides to assist the participants (both experienced and inexperienced users) to properly evaluate the usability of \projectname in the real-life smart home.

%% file: relatedwork.tex
\section{Related Work} \label{sec:related}

\begin{table}[t!]
\centering
\resizebox{\columnwidth}{!}{
\setlength{\tabcolsep}{0.01in}
\renewcommand{\arraystretch}{1}
{\scriptsize{
\begin{tabular}{|lccccccc|}
\hline
\textbf{\begin{tabular}[c]{@{}c@{}}Prior \\Work\end{tabular}} &
\textbf{\begin{tabular}[c]{@{}c@{}}Domain\end{tabular}} & \textbf{\begin{tabular}[c]{@{}c@{}}Multi-user\\Multi-device \\environment\end{tabular}} &
\textbf{\begin{tabular}[c]{@{}c@{}}A.C. \\Threat \\ Model\end{tabular}} &
\textbf{\begin{tabular}[c]{@{}c@{}}User\\Interface\end{tabular}} &
\textbf{\begin{tabular}[c]{@{}c@{}}Conflict\\resolution\end{tabular}} &
\textbf{\begin{tabular}[c]{@{}c@{}}Overhead\\analysis\end{tabular}} &
\textbf{\begin{tabular}[c]{@{}c@{}}A. C.\\Language\end{tabular}} \\ \hline
 \hline
\rowcolor{light-gray}
xShare~\cite{liu2009xshare} & Smartphone  & \wcircle & \wcircle  & \bcircle  & \wcircle  &  \bcircle &    \wcircle    \\ 
DiffUser~\cite{ni2009diffuser} & Smartphone  & \wcircle  & \wcircle & \bcircle  & \wcircle  &  \bcircle &    \wcircle    \\
\rowcolor{light-gray}
Capability-based A. C.~\cite{GUSMEROLI20131189} & IoT network  & \bcircle  &    \wcircle & \bcircle  & \wcircle &  \wcircle & \wcircle      \\
Situation-based A. C.~\cite{Schuster}   & Smart home  & \bcircle   &   \wcircle  & \wcircle  & \wcircle &  \wcircle  & \bcircle    \\
\rowcolor{light-gray}
Expat~\cite{yahyazadeh2019expat} & Smart home  & \bcircle &   \wcircle  & \wcircle  & \wcircle &  \bcircle    & \bcircle        \\
Zeng et al.~\cite{zeng2019understanding}   & Smart home  & \bcircle &  \wcircle & \bcircle  & \wcircle &  \wcircle & \wcircle        \\
\rowcolor{light-gray}
\projectname & Smart home  & \bcircle  & \bcircle  & \bcircle &  \bcircle    & \bcircle  &   \bcircle      \\
\hline
\end{tabular}
}}}
\caption{Comparison between \projectname and other access control mechanisms (A.C. stands for Access Control).}
\vspace{-0.4in}
\label{compare}
\end{table}

Rather than providing fine-grained user access control, most of the prior works emphasize on limiting malicious activities via controlling app access~\cite{demetriou2017hanguard, fernandes2016flowfence}. Moreover, several works focus on device access control and authentication on an IoT network for single-user scenarios~\cite{6915840, Agadakos:2016:LAU:2991079.2991090, 7581588, Jacobs, rajtmajer2017ultimatum}. In a recent work, He et al. present a detailed smart home user study that portrays users' concerns of fine-grained access control in multi-user smart environments~\cite{217501}. Here, authors conducted the user study among 425 smart home users and outlines access control needs based on users preferences, social norms, and mutual relationships. Zeng et al. discuss their findings related to security and privacy concerns among smart home users~\cite{205156}. Authors selected 15 smart home users and outlined their security and privacy concerns and summarizes users' actions in security-affected scenarios. In both works, smart home users raise their concerns regarding the need of access control mechanism in SHS. In addition, these studies also summarize design specifications to reflect users' needs in an access control mechanism. Matthews et al., also points out relevant issues with smart home users that share the same devices and accounts~\cite{Matthews:2016:SJG:2858036.2858051}. However, no explicit solution for multi-user access is proposed in any of these works.

In other works, researchers explore different access control strategies when multiple users share a single IoT device. Liu et al. suggested a user access framework for the mobile phone ecosystem called \textit{xShare}, which provides policy enforcement on file level accesses~\cite{liu2009xshare}. Ni et al. presented \textit{DiffUser}, a user access control model for the Android environment based on access privileges~\cite{ni2009diffuser}, which is only effective for a single device. Tyagi et al. discussed several design specification needed for multi-party access control in a shared environment~\cite{tyagi2016depth}. Aside from these works, there are few prior works proposing access control systems for multi-user multi-device SHS. Gusmeroli et al. suggested a capability-based access control for users in a multi-device environment~\cite{GUSMEROLI20131189}. However, this system is not flexible enough to express the real needs of the users. Jang et al. presented a set of design specification for access control mechanism based on different use scenarios of multi-user SHS~\cite{Jang:2017:EMC:3139937.3139941}. Schuster et al. proposed a situation-based access control in the SHS which considers different environmental parameters~\cite{Schuster}. Here, the authors considered state of the device along with the location of the users to determine a valid access request. However, this work does not solve the conflicting demands of multiple users. Yahyazadeh et al. presented \textit{Expat}, a policy language to define policies based on user demands~\cite{yahyazadeh2019expat}. In a recent work, Zeng et al. built an access control prototype with different access control options for smart home users~\cite{zeng2019understanding}. Here, the authors considered four different access control mechanisms and assessed in a month-long user study among seven households to understand the users' needs and improve the design. Although authors built a proof-of-concept framework to perform a detailed user study and outlined the access control needs in smart home, they did not implement the framework in real-life systems and did not consider user conflicts while operating in a multi-user smart environment. 

\vspace{1pt}\noindent \textbf{Differences from existing works.} \projectname was built upon considering prior user studies~\cite{217501, zeng2019understanding}. \projectname presents an access control system designed for multi-device multi-user smart home systems that provides a fine-grained access control to the users considering (1) easy new user addition with priority levels, (2) device restrictions for specific users, (3) automatic policy negotiation for conflicts, (4) easy policy assignment for multiple users, (5) different threats arising from over-privileged users, (6) Real-life implementation in smart home platform, (7) Effectiveness evaluation with real-life users, and (8) minimum overhead in real-life deployment. Table~\ref{compare} summarizes the differences of \projectname from other existing solutions. 
\vspace{-0.1in}

%% file: conclusion.tex
\vspace{-0.1in}
\section{Conclusion} 
\label{sec:conclusion}

In smart home systems, multiple users have access to multiple devices simultaneously. In these settings, users may want to control and configure the devices with different preferences which give rise to complex and conflicting demands. In this paper, we proposed \projectname, an access control system that addresses the diverse and conflicting demands of different users in a shared multi-user smart home system. \projectname implements a priority-based policy negotiation technique to resolve conflicting user demands in a shared smart home system. We implemented \projectname on real settings with multiple users and evaluated its performance via real devices. \projectname successfully covers the users' needs, and our extensive evaluations showed that \projectname is effective in resolving the conflicting requests and enforcing the policies without significant overhead. Also, we tested \projectname against five different threats and found that \projectname effectively identifies the threats with high accuracy. 

%% file: appendix1.tex
\appendix{}
\label{sec:appendix}

\setcounter{figure}{0}
\setcounter{table}{0}
\setcounter{lstlisting}{0}
\setcounter{section}{0}
\setcounter{equation}{0}

\section{Policy Negotiation Algorithm} \label{policyalgorithm}

During policy negotiation, each policy clause is compiled into a quintuple, $\Psi =  \{P,U,D,\mathcal{C},A\}$, where $P$ is the policy assigner (that shows who states this clause), $U$ is the assignee (about whom this statement is), $D$ is the targeted smart device,  $\mathcal{C}$ is a set of conditions over $D$ and $U$, and configurable environmental attributes, and finally $\mathcal{A}\in\{demand,restrict\}$ is the action requested by this statement when the set of conditions are satisfied. \projectname implements an algorithm to solve the policy conflicts represented as follows:

\begin{equation}
\footnotesize
interfere(\Psi_i,\Psi_j) \gets  U_i = U_j \land D_i = D_j
\end{equation}
\begin{equation}
\footnotesize
\begin{split}
\begin{aligned}
hard\_conflict(\Psi_i,\Psi_j) \gets interfere(\Psi_i,\Psi_j) \land ( \\
    (A_i \neq A_j  ~\land \forall c \in \mathcal{C}_i \cap \mathcal{C}_j :  \Theta(\mathcal{V}(c,\mathcal{C}_i), \mathcal{V}(c,\mathcal{C}_j) ) ) \\
    \lor ( A_i = A_j  ~\land \exists c \in \mathcal{C}_i \cap \mathcal{C}_j :  \neg \Theta(\mathcal{V}(c,\mathcal{C}_i), \mathcal{V}(c,\mathcal{C}_j) ) )   
) \\ 
\end{aligned}
\end{split}
\end{equation}
\begin{equation}
\footnotesize
\begin{split}
\begin{aligned}
soft\_conflict(\Psi_i,\Psi_j) \gets interfere(\Psi_i,\Psi_j) \land ( \\
    (A_i = A_j  ~\land \forall c \in \mathcal{C}_i \cap \mathcal{C}_j :  \Theta(\mathcal{V}(c,\mathcal{C}_i), \mathcal{V}(c,\mathcal{C}_j) ) ) \\ 
    \lor ( A_i \neq  A_j  ~\land \exists c \in \mathcal{C}_i \cap \mathcal{C}_j : \mathcal{V}(c,\mathcal{C}_i) \neq \mathcal{V}(c,\mathcal{C}_j) )   
) 
\end{aligned}
\end{split}
\end{equation}

\begin{equation}
\footnotesize
HPC(\Psi_i,   \Psi_j)  \gets  hard\_conflict(\Psi_i,\Psi_j) \land \Xi(P_i) \neq \Xi(P_j) 
\end{equation}
\begin{equation}
\footnotesize
SPC  (\Psi_i,   \Psi_j)   \gets  soft\_conflict(\Psi_i,\Psi_j) \land \Xi(P_i) \neq \Xi(P_j)
\end{equation}
\begin{equation}
\footnotesize
HCC  (\Psi_i,    \Psi_j)   \gets  hard\_conflict(\Psi_i,\Psi_j) \land \Xi(P_i) = \Xi(P_j)
\end{equation}
\begin{equation}
\footnotesize
SCC  (\Psi_i,    \Psi_j)   \gets  soft\_conflict(\Psi_i,\Psi_j) \land \Xi(P_i) = \Xi(P_j)
\end{equation}
\begin{equation}
\footnotesize
\begin{split}
\begin{aligned}
RC  (\Psi_i, \psi_j)   \gets  Restriction\_conflict(\Psi_i, \psi_j) \land \Xi(P_i) > \Xi(P_j) \\\land A_i = restrict
\end{aligned}
\end{split}
\end{equation}

\noindent where $\Psi_i,\Psi_j$ is the evaluated pair of policies, and  $\mathcal{V}(c,C)$ is the value function that returns the value of conditional $c$ in the set $C$, $\Theta(x,y)$ checks the overlap between the provided $(x,y)$ tuple and $\Xi(u)$ returns the priority of user $u$ as the value of user's assigned priority class.
\section{Policy Negotiation Process} \label{negotiation}

The negotiation $\mathcal{N}$ between two given policy clauses $(\Psi_i,\Psi_j)$ can be formally expressed and computed by Equation~\ref{eq:nagotiation}. 
\begin{equation}
\scriptsize
 \centering
        \mathcal{N}(\Psi_i,\Psi_j)= 
\begin{cases}
 \centering
    \begin{cases} 
        \Psi_i & \text{if } \Xi(P_i) > \Xi(P_j) \\
         \Psi_j & \text{otherwise}
    \end{cases},&   \text{if } HPC(\Psi_i,\Psi_j) \\
    \begin{cases} 
        \{P_i \cup P_j,U_i,D_i,C_i \cup C_j,A_i\} & \text{if } A_i = A_j \\
         \{P_i \cup P_j,U_i,D_i, C_i \cup \neg C_j,A_i\} & \text{otherwise}
    \end{cases},&   \text{if } SPC(\Psi_i,\Psi_j) \\
    \begin{cases} 
        majority\_vote(\Psi_i,\Psi_j)  & \text{if } binary(D_i) \\
        arbitrate(\Psi_i,\Psi_j) & \text{otherwise}
    \end{cases},&   \text{if } HCC(\Psi_i,\Psi_j) \\
    \begin{cases} 
        \{P_i \cup P_j,U_i,D_i, C_i \cup C_j,A_i\} & \text{if } A_i = A_j \\
         \{P_i \cup P_j,U_i,D_i, C_i \cup \neg C_j,A_i\} & \text{otherwise}
    \end{cases},&   \text{if } SCC(\Psi_i,\Psi_j) \\
\end{cases} 
\label{eq:nagotiation}
\end{equation}

In the case of a hard priority conflict (HPC), (e.g., mother vs. child with contradicting clauses) \projectname prioritizes the clause of the user with the higher priority (e.g., mother). For hard competition conflict (HCC), both users with overlapping conditions are notified and \projectname offers a common operating condition to both. This common condition is enforced as a policy to the device upon users' agreement. On the other hand, in the case of both soft priority (SPC) and soft competition conflicts (SCC), the result of the negotiation is a new clause with common set of conditions. For restriction conflict, both restricted user and policy assigner are notified and if the policy satisfies conditions in Equation~\ref{eq:nagotiation}, the restriction policy is enforced in the device.

\section{Devices Used During Evaluation}\label{devicelst}
We present a detailed list of devices used during the implementation and evaluation of \projectname in Table~\ref{tab:devicelist}. 

\begin{table}[h!]
\renewcommand{\arraystretch}{0.82}
\centering
\fontsize{8}{10}\selectfont
\resizebox{0.45\textwidth}{!}{
\begin{tabular}{|p{2.3cm}|p{4.1cm}|c|}
\hline
\textbf{\begin{tabular}[c]{@{}c@{}}Device Type\end{tabular}} &  \textbf{Model} & \textbf{\begin{tabular}[c]{@{}c@{}}Quantity\end{tabular}} \\ \hline\hline
Smart Home Hub & Samsung SamrtThings Hub & 1 \\\hline
Smart Light & Philips Hue Light Bulb & 4\\\hline
Smart Lock & Yale B1L Lock with Z-Wave Push Button Deadbolt & 1\\\hline
Smart Camera & Arlo by NETGEAR Security System & 1 \\\hline
Smart Thermostat & Ecobee 4 Smart Thermostat & 1 \\\hline
Motion Sensor & Fibaro FGMS-001 ZW5 Motion Sensor with Z-Wave Plus Multisensor  & 6 \\\hline
Temperature Sensor & Fibaro FGMS-001 ZW5 Motion Sensor with Z-Wave Plus Multisensor & 1 \\\hline
Door Sensor & Samsung Multipurpose Sensor & 2\\ \hline
\end{tabular}}

\caption{Devices and sensors used in our smart home setup to evaluate \projectname.}
\vspace{-0.3in}
\label{tab:devicelist}
\end{table}
\vspace{-0.1cm}
\section{\textsc{KRATOS}-enabled SmartThings App}\label{patch}
We provide an example of \projectname-enabled SmartThings App.
\begin{lstlisting}[caption=Policy enforced at install-time, label=App Patching, belowskip=0.1\baselineskip]
definition(
    name: "Big Turn ON modified",
    namespace: "smartthings",
    author: "Anonymous",
    description: "Turn your lights on when the SmartApp is tapped.",
    category: "Convenience",
    iconUrl: "https://s3.amazonaws.com/smartapp-icons/Meta/light_outlet.png",
    iconX2Url: "https://s3.amazonaws.com/smartappicons/Meta/light@2x.png"
)
import groovy.time.*
preferences {
	section("When I touch the app, turn on...") {
	input "switches", "capability.switch", multiple: false
        input name: "email", type: "email", title: "Email", description: "Enter Email Address", required: true, displayDuringSetup: true}}
def installed()
{       atomicState.SmartLightTimes = [:]
   	atomicState.SmartLightAdmins = [:]
   	atomicState.SmartLightUsers = [:]
   	atomicState.SmartLightDevID = [:]
   	atomicState.SmartLightTimeStart = [:]
   	atomicState.SmartLightTimeEnd = [:]
    
    log.debug "${new Date()}"
  	getSmartLightJsonData()
    
    def item = atomicState.SmartLightUsers.indexOf(email)
    if (item>=0){
    	int index = atomicState.SmartLightUsers.indexOf(email)
        def between = timeBetween (atomicState.SmartLightTimeStart[index], atomicState.SmartLightTimeEnd[index])
        if (between == true){
            subscribe(location, changedLocationMode)
            subscribe(app, appTouch)
            log.info app.getAccountId()}}
}
def updated()
{   atomicState.SmartLightTimes = [:]
   	atomicState.SmartLightAdmins = [:]
   	atomicState.SmartLightUsers = [:]
   	atomicState.SmartLightDevID = [:]
   	atomicState.SmartLightTimeStart = [:]
   	atomicState.SmartLightTimeEnd = [:]
    getSmartLightJsonData()
    
    def item = atomicState.SmartLightUsers.indexOf(email)
    if (item>=0){
    	int index = atomicState.SmartLightUsers.indexOf(email)
        def between = timeBetween (atomicState.SmartLightTimeStart[index], atomicState.SmartLightTimeEnd[index])
        if (between == true){
            unsubscribe()
            subscribe(location, changedLocationMode)
            subscribe(app, appTouch)}}
}
def changedLocationMode(evt) {
	log.debug "changedLocationMode: $evt"
	switches?.on()}
def appTouch(evt) {
	log.debug "appTouch: $evt"
	switches?.on()}
def getSmartLightJsonData(){
    def listTimes = []
    def listAdmins = []
    def listUsers = []
    def listIDs = []
    def listTimeStarts = []
    def listTimeEnds = []
    def params = [uri: "https://mywebserver/xxxyyyzzz/2/public/values?alt=json",]
    try {
        httpGet(params) { resp ->
            for (object in resp.data.feed.entry){
				listTimes.add (object.gsx$time.$t)
                listAdmins.add (object.gsx$adminemail.$t)
                listUsers.add (object.gsx$restricteduseremail.$t)
                listIDs.add (object.gsx$deviceid.$t)  
                listTimeStarts.add (object.gsx$timerangestart.$t)    
                listTimeEnds.add (object.gsx$timerangeend.$t)    
            }
            atomicState.SmartLightTimes = (listTimes)
            atomicState.SmartLightAdmins = (listAdmins)
            atomicState.SmartLightUsers = (listUsers)
            atomicState.SmartLightDevID = (listIDs)
            atomicState.SmartLightTimeStart = (listTimeStarts)
            atomicState.SmartLightTimeEnd = (listTimeEnds)}
    } catch (e) {
        log.error "something went wrong: $e"}
}

def timeBetween(String start, String end){
    long timeDiff
    def now = new Date()
    def timeStart = Date.parse("yyy-MM-dd'T'HH:mm:ss","${start}".replace(".000-0400",""))
    def timeEnd = Date.parse("yyy-MM-dd'T'HH:mm:ss","${end}".replace(".000-0400",""))
    long unxNow = now.getTime()
    long unxEnd = timeEnd.getTime()
    long unxStart = timeStart.getTime()
    if (unxNow >= unxStart && unxNow <= unxEnd)
    	return true
    else
    	return false
 }
\end{lstlisting}